%% file: main.tex
\documentclass{article}
\usepackage{arxiv}

 \usepackage{booktabs}
 \usepackage{graphicx}
\usepackage{natbib}
\usepackage{hyperref}

\newcommand{\ReQ}[1]{{\textit{#1}}}

\usepackage{soul}
\usepackage{float}
\usepackage{pdflscape}
\usepackage{longtable}
\usepackage{dirtytalk}

\usepackage[utf8]{inputenc} 
\usepackage[T1]{fontenc}    
\usepackage{hyperref}       
\usepackage{url}            
\usepackage{booktabs}       
\usepackage{amsfonts}       
\usepackage{nicefrac}       
\usepackage{microtype}      
\usepackage{cleveref}       
\usepackage{lipsum}         
\usepackage{graphicx}
\usepackage{natbib}
\usepackage{doi}

\usepackage{epstopdf}

\renewcommand{\shorttitle}{Automating the Identification of High-Value Datasets in Open Government Data Portals}
\renewcommand{\headeright}{}
\renewcommand{\undertitle}{}

\title{Automating the Identification of High-Value Datasets in Open Government Data Portals: A US Municipalities Case Study}

\author{ \href{https://orcid.org/0000-0002-1801-3403}{\includegraphics[scale=0.06]{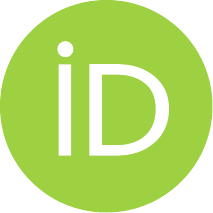}\hspace{1mm}Alfonso~Quarati}\thanks{Coresponding Author} \\
	Institute for Applied Mathematics and Information Technology\\
	National Research Council\\
	Genoa,Italy \\
	\texttt{alfonso.quarati@cnr.it} \\
	\And
	\href{https://orcid.org/0000-0002-0532-3488}{\includegraphics[scale=0.06]{orcid.pdf}\hspace{1mm}Anastasija~Nikiforova} \\
	Institute of Computer Science\\
	Faculty of Science and Technology, University of Tartu\\
	Tartu,Estonia\\
	\texttt{nikiforova.anastasija@gmail.com} \\
}

\begin{document}
\maketitle

\begin{abstract}
Recognized for fostering innovation and transparency, driving economic growth, enhancing public services, supporting research, empowering citizens, and promoting environmental sustainability, High-Value Datasets (HVD) play a crucial role in the broader Open Government Data (OGD) movement. However, identifying HVD presents a resource-intensive and complex challenge due to the nuanced nature of data value. Our proposal aims to automate the identification of HVDs on OGD portals using a quantitative approach based on a detailed analysis of user interest derived from data usage statistics, thereby minimizing the need for human intervention. The proposed method involves extracting download data, analyzing metrics to identify high-value categories, and comparing HVD datasets across different portals.  
This automated process provides valuable insights into trends in dataset usage, reflecting citizens' needs and preferences. The effectiveness of our approach is demonstrated through its application to a sample of US OGD city portals.
 The practical implications of this study include contributing to the understanding of HVD at both local and national levels. By providing a systematic and efficient means of identifying HVD, our approach aims to inform open governance initiatives and practices, aiding OGD portal managers and public authorities in their efforts to optimize data dissemination and utilization.
\end{abstract}




\section{Introduction}

Open government data (OGD) refers to the movement of making government-held data freely available to the public in machine-readable formats. Over the last decades, this movement has gained global traction, with most nations participating by making their data accessible to a wide audience, including civil society, businesses, academia and research, developers, journalists, non-governmental organizations, and public sector agencies, which serve as both data publishers and data consumer \citep{ubaldi2013open,kassen2021understanding,schwoerer2022whose,quaratiJIS2020}. 
 However, the mere availability of vast amounts of data spanning areas such as economics, healthcare, transportation, education, and environmental data—has shown limited impact without active and meaningful utilization, where the true value and impact of OGD lies in the public's interest and the likelihood of data reuse by various stakeholders, transforming raw data into tangible public value encompassing social, economic, and environmental benefits \citep{ubaldi2013open,european2020impact,page2023open}. As such, the focus switches to releasing high-value datasets, where "quality over quantity" is prioritized (quality in this context refers to the value of data). 
 
 Many governments around the globe recognize the importance of high-value data in their ability to foster transparency, drive economic growth, enhance public services, support research, empower citizens, and promote environmental sustainability, and make attempts to identify these data \citep{ubaldi2013open,gasco2018promoting,gao2023understanding}. By leveraging these datasets, governments, businesses, researchers, and citizens can collaborate to create a more informed, efficient, and equitable society.

The identification and release of High-Value Datasets (HVD) play a pivotal role in the broader OGD movement, aiming to enhance the accessibility and impact of government data. However, identifying datasets that can be seen as HVD presents a complex challenge due to the nuanced nature of data value. This determination necessitates considering perspectives from both data publishers and users, which often diverge even within the same category \citep{ubaldi2013open,european2020impact}, e.g., users representing civil society and business can have different interests. While many countries, particularly those within the European Union, adhere to pre-defined categories of HVD outlined by the Open Data Directive \citep{HVD-EU}, there's a growing imperative to expand this classification. There is a call to identify additional datasets recognized as HVD that can be considered to be such within a certain geographical area such as countries, municipalities, or regions \citep{european2020impact,Utamachant,Nikiforova23,van2022trends,Nikiforova2021}. This localized approach ensures that datasets deemed high-value resonate closely with the needs and priorities of local stakeholders, fostering greater relevance and utility.

Identifying HVD within the governmental domain requires tackling several critical questions regarding the classification criteria and the selection of suitable indicators and measurement methodologies, preferably (semi-)automatic \citep{Nikiforova23,autoHVD}.

The most progress in this regard can be seen to be made by the European Commission through the Open Data Directive \citep{PSI,HVD-EU,HVD-EU2,HVD-EU2}. This directive not only outlines categories of datasets designated as HVD but also identifies specific datasets falling into these categories, along with requirements for their publication or annotation on portals, as well as suggests a guide for EC Member States in discovering HVD on their portals \citep{autoHVD}. However, the complexity of the task has led to a partly semi-automated approach with a predominance of manual activities surrounding the process. Consequently, this method consumes considerable resources. Moreover, considering the focus of this report, all advancements concern datasets and categories designated as HVD by the European Commission, rather than those identified by individual Member States. 

To address these issues, this study aims to develop an automated approach to identify High-Value Data based on the actual interest shown by users of OGD portals, by analyzing data usage statistics obtained programmatically (i.e. via APIs). 
While we recognize that mere downloading does not guarantee subsequent dataset reuse \citep{Nikiforova23}, and that true value lies in effective utilization, aligning with both the research and practice, \textit{download statistics} serve as a SMART (Specific-Measurable-Achievable-Relevant-Ttime-bound) and \textit{ex-post} indicator, contributing to the identification of HVD, which combined with other methods can ensure a more comprehensive approach to HVD determination.

As such, our approach comprises three key steps. Firstly, we programmatically extract download data from the metadata of portal datasets. Subsequently, we analyze these data using various metrics to identify the most frequently accessed dataset categories. This analysis includes the development of a specialized metric called the \textit{High-Value Data Index}, which evaluates 'high-value datasets categories' by leveraging download data in conjunction with other factors such as the total number of datasets within each category. 
Finally, we compare the identified HVD categories across portals representing a selected area - country, region - highlighting the most prevalent categories among a selection of portals. This comparison involves aligning portal categories to standardize the analysis, facilitating meaningful comparisons despite variations in denotations and category definitions across portals by drawing on the methodology developed by Pinto et al. \citep{Pinto20}.

We present the suggested approach, focusing on a sample of major U.S. metropolitan statistical areas. Recognizing the unique characteristics and scale of the country, this analysis aims to pinpoint categories of datasets of high value within specific U.S. regions. This serves a dual purpose: firstly, identifying topics that hold national significance for the United States; and secondly, demonstrating the effectiveness of our approach in action, thereby providing a blueprint for its future use.

As such, the study makes theoretical contributions by developing the High-Value Data index (HVDi), a tool designed to assess dataset categorization across portals. This HVDi enables effective comparisons and identification of HVD within specific areas. Additionally, the research introduces a novel methodology for automatically analyzing downloaded data from OGD portals to identify high-value data categories. This innovative approach enriches the theoretical framework by providing region-specific insights essential for effective data prioritization and governance.

On the practical side, the study offers a structured methodology for identifying data with high reuse potential, exemplified by the case of the U.S. This blueprint not only aids in data prioritization but also supports open governance initiatives. By highlighting disparities and commonalities in dataset utilization, the research enhances localized understanding of HVD, informing open governance efforts to improve data homogeneity and interoperability within an area (country, region). Moreover, the practical methodology for identifying and comparing HVD across different portals provides actionable insights by automatically generating visualizations of the results, which improves the perception of results. These insights guide the prioritization of datasets crucial for open governance and data reuse efforts.

The article is structured as follows: the following section provides an overview of the HVD determination approaches in government initiatives and research. Section 3 defines the research questions, and presents the methodology, sample selection, and data collection method. Section 4 presents the results, Section 5 presents the Discussion, Limitations, and Implications for theory and practice. Finally, Section 6 concludes the paper.

\section{Background}
In this section, we briefly discuss the current state of HVD in individual country contexts referring to the research, and practical developments.

\subsection{Government and local initiatives}

The identification and publication of HVD are critical components in maximizing the utility of open government data and the general success of OGD as an initiative. To this end, many governments - national and local - have put efforts into identifying and publishing high-value data. However, whereas there is some clear evidence of these efforts as we describe below, it must be acknowledged that some efforts of making an attempt to open HVDs or even succeeding in it remain unknown being hidden from the public (also in line with \citep{Nikiforova23}). This, in turn, leads to a lack of standardization of this lifecycle and as such slows the adoption of respective techniques. 
Let us now refer to some known efforts towards HVD opening as a result of our desk research.

HVD identification and their opening is part of the Victorian community - a state in southeastern \textbf{Australia} - \textit{DataVic} access policy guidelines, according to which HVD refers to “\textit{a dataset that is likely to be of interest to the Victorian community, and/or a dataset that has potential for valuable reuse}”. As such, they defined a set of criteria for determining HVD, according to which a dataset is considered ‘high value’ if it: (1) is central to the department/agency functions, (2) has been requested by users via ‘\textit{suggest a dataset}’ feature, (3) has previously/regularly been provided under the \textit{Freedom of Information Act 1982 8}, (4) supports a major reporting process of government, (5) planning data, (6) spatial data, (7) transport data, (8) administrative data, (9) financial data, as well as other data determined as HVD by agencies, which are expected to examine their data. 

Another example comes from \textit{Indian Urban Exchange (IUDX)} – initiated and funded by the Ministry of Housing and Urban Affairs and supported by the Ministry of Electronics and Information Technology and National Institution for Transforming India (NITI) Aayog of \textbf{India} - to provide a data exchange platform to Indian cities. Collaborating with over 30 Smart cities in India, IUDX identified over 225 HVD across various domains within Urban Governance (as of the end of 2020). HVD are referred in their case to datasets that are “\textit{instrumental in deriving important benefits for the society, the environment and the economy, in particular, because of their ability towards creating value-added services for efficiency and convenience by making the best use of AI/ML technologies}.” To this end, datasets and data owners were identified supplying them with the examples referred to as “value” in the form of potential use-cases. Identified HVDs span 15 primary domains – mobility, traffic, solid waste management, environment, water distribution, energy and utilities, healthcare, emergency services, tourism, revenue collection, citizen grievances, smart elements, street light, citizen security, GIS, with 8 more domains found of (secondary) importance, namely, agriculture, crime and justice, earth observations, education, finance and contracts, government, science and research, social welfare. For all 23 domains a list of HVD categories is identified, defining what is called in this case “\textit{HVD exercise}” that refers to the complete process of identifying the HVDs, carrying out their valuation, assessing their quality, creating value, identifying the right monetization approach, and eventually monetizing it for data owners.
 As of May 2024, data.gov.in catalog of the  \textit{Open Government Data (OGD) Platform India} has  557 HVDs.

Similarly to the Indian case, i.e., with the focus on local governments and their data opening, in 2016, \textit{data.overheid.nl} (the \textbf{Netherlands}), a Municipal High-Value List has been compiled by the national government in collaboration with municipalities, the Digital Urban Agenda, and VNG/KING, intended
to assist municipalities in opening data and prioritizing specific datasets \citep{overheidspublicaties_relevant_nodate}. These HVDs fall under the category of '\textit{Data with Impact},' which in addition to HVD consists of “\textit{reference datasets}” and “\textit{nationwide datasets}”, including not only municipal HVDs but also local datasets. The value of a dataset and its classification as an HVD is determined by several factors, namely (1) transparency, (2) legal obligations, (3) cost reduction, (4) target audience, and (5) potential for reuse. To identify potentially high-value datasets, the G8 Open Data Charter and its fourteen categories of datasets were used as a reference, along with the Open Data Barometer (by the Open Data Institute), the Open Data Maturity Report, the Global Open Data Index (by the Open Knowledge Foundation), Country Factsheets from the Government at a Glance OECD, and Dutch benchmarks from 2014 to 2016, serving as a guide for data owners to open the most relevant and potentially useful datasets.

In \textbf{Canada}, the effort to identify HVDs has been ongoing for several years, with each government jurisdiction developing its criteria for this purpose. The Canada Open Government Working Group (COGWG) was later assigned the task—part of the Third Biennial Plan to the Open Government Partnership (2016-18) Commitment\#16 to \say{develop a list of high-value, priority datasets for release in collaboration with key jurisdictions to make it easier for Canadians to compare data across different governments} \citep{government_of_canada_third_2002}. In 2018, they came up with a set of criteria and corresponding examples of common dataset types derived from a jurisdictional review of HVD criteria, surveys, and the International Open Data Charter (ODC). The identified criteria encompassed five key aspects: (1) identification of social, environmental, and economic conditions as HVD should assist in identifying and understanding them within the community or nation; (2) enhancement of public services outcomes - HVD should contribute to improving outcomes in public services, ultimately leading to better service delivery and citizen satisfaction; (3) promotion of innovation and sustainable economic growth - HVD should foster innovation and support sustainable economic growth by providing valuable insights for businesses, researchers, and policymakers; (4) increased transparency and accountability - HVD release should enhance transparency, accountability, and the flow of information within the government, empowering citizens to engage with government data and processes; (5) high demand from the community - HVD should be in high demand from the community, aligning with the needs and values of citizens. This demand can be gauged through requests for specific datasets by citizens and communities, serving as an indicator of public interest and demand for these data. Based on this set of criteria, 17 HVDs were determined for consideration across federal, provincial, and territorial governments. However, these datasets were not mandatory for governments to release, suggesting this list to be considered in prioritizing datasets for release. The longer-term objective of the working group is the standardization of HVD dissemination.
 While there was considered to be limited progress over the years, according to the current National Action Plan (2022-2024) on Open Government Commitment, \say{Collecting and releasing high-value data related to various policing activities, workforce composition, and more} is one of the milestones and respective indicators that Royal Canadian Mounted Police is responsible for within it. Although its deadline was May 2023, the last update (as of the end of 2023) announced the delay of the annual report on Police Disclosure of Information due to internal resource shortages and technical issues with the promised release scheduled for January 2024 with no further updates.

A similar approach is followed by \textbf{Taiwan}, which, is part of the Taiwan Open Government National Action Plan 2021-2024, where prioritization and opening of HVD is part of its first commitment as part of "Completing Government Open Data and Data Sharing Mechanism", which, in turn, is the first step to "Promote Open Data and Freedom of Information" seeking for a more comprehensive system and mechanism to optimize the value of government open data usage. This commitment entails prioritizing HVD opening, enhancing data standards and format quality, and establishing processes to meet public data needs. The plan identifies six key areas of HVDs: (1) climate and environment, (2) disaster prevention and relief, (3) transportation and transit, (4) healthcare and medical services, (5) energy management, and (6) social assistance. Specific datasets within these categories have been identified and assigned to respective Taiwanese agencies for opening, whereas responsible bodies are then using their approaches towards prioritization of HVD, such as relying on the G8 Open Data Charter, Global Data Barometer, EC OD Directive, Australia, USA, with the EU OD Directive being the most popular choice among responsible ministries \citep{tseng2024unlocking}.

To sum up, today several governments of both national and local levels have taken initiatives to prioritize HVD opening. They predominantly follow two approaches: (1) ex-ante and (2) ex-post, where the former involves the identification of datasets to be opened that are or may be of interest to open data end-users, while the latter - ex-post - focuses on datasets that are already available, either (2.1) assessing the impact and current interest of existing datasets with the further decision on their maintenance, or (2.2) considering the patterns observed in user interest and/or impact of those datasets, determining topics for which more datasets should be opened. 

In the pursuit of identifying HVD, government and local initiatives employ different methodologies, which are often mixed considering the complexity and multi-faceted nature of data value. These methodologies can be categorized into several groups based on their underlying principles, where one group revolves around \textit{economic and market-based approaches}. Here, data is viewed as a valuable asset to be monetized directly through trading or by creating services around it. Market value is determined by assessing the prices of similar datasets in the market, while \textit{cost-based valuation} considers the expenses incurred in producing and maintaining the dataset. Additionally, \textit{game-theory-based} approaches employ strategic analysis to estimate the data worth within competitive landscapes. Another group centers on \textit{usage and contribution-based methodologies}. These approaches gauge the dataset's value based on user engagement and contributions, indicating its relevance and utility. \textit{Download statistics} are often utilized to measure user demand and infer the data value. \textit{Strategic and utility-based approaches} form the third category, where the dataset's potential to drive key initiatives for cost reduction or revenue enhancement is evaluated. Relative value assessments compare datasets against one another to determine their importance and priority.
 Lastly, \textit{criteria-based methodologies} evaluate datasets based on specific criteria such as \textit{transparency}, \textit{legal obligations}, and \textit{potential for reuse}. 

However, while the above examples reflect the national and local government efforts, there is another example worth discussing that represents perhaps the most impressive efforts to date to identify and open up HVDs, namely, the Open Data Directive (formerly Public Sector Information Directive (PSI Directive)). This is also supported by the Taiwanese example, where the Open Data Directive has gained the most popularity among ministries responsible for HVD opening, using it as the reference model.

Despite some national efforts, the HVD opening movement in the EU can be considered to be launched in 2019, when the term “high-value dataset” was adopted by the European Parliament and the Council of the European Union introduced the term initiating efforts to identify and promote HVDs within EU Member States. The European Commission, through the Open Data Directive (Directive), delineates HVD as “\textit{datasets holding the potential to (i) generate significant socio-economic or environmental benefits and innovative services, (ii) benefit a high number of users, in particular SMEs, (iii) assist in generating revenues, and (iv) be combined with other datasets}”.
Directive (EU) 2019/1024 of the European Parliament and of the Council of 20 June 2019 defined six thematic data categories of HVDs: \textit{geospatial}, \textit{earth observation and environment}, \textit{meteorological}, \textit{statistics}, \textit{companies and company ownership}, and \textit{mobility}. These categories were subsequently detailed in Commission Implementing Regulation (EU) 2023/138, which Member States must adhere to by June 2024. In 2023, "Identification of data themes for the extensions of public sector High-Value Datasets" \citep{HVD-EU2} was published, proposing seven additional categories to be potentially added acknowledged as HVDs, namely: \textit{climate loss}, \textit{energy}, \textit{finance}, \textit{government and public administration}, \textit{health}, \textit{justice and legal affairs}, \textit{linguistic data}. This regulation aims at enhancing harmonization and interoperability by defining specific datasets, their granularity, key attributes, geographical coverage, and requirements for reuse, including licensing, format specifications, update frequency, accessibility, and metadata standards, while ensuring readiness for reuse and compliance with predefined macro-characteristics. These characteristics include economic, environmental, and social benefits, facilitation of innovative services, and adaptation to advanced digital technologies like artificial intelligence, distributed ledger technologies (DLT), and the Internet of Things (IoT). Additionally, the Directive emphasizes the importance of reuse and the support of public authorities in fulfilling their mandates \citep{european2020impact}. 
The recent Open Data Maturity Report (ODM) \citep{page2023open} indicates active preparations among Member States to open HVDs, with 25 Member States working towards implementing the Commission Implementing Regulation (EU) 2023/138. Certain countries, such as Estonia, Finland, Denmark, Latvia, Czechia, and Slovenia, are notably ahead in their preparations. However, while progress is observed in categories like geospatial and statistics datasets, other HVDs are being opened up less actively. From a lifecycle perspective, tasks related to identification, inventory, and legal barriers are more advanced compared to meeting technical requirements such as metadata quality and standardized structures.

As a follow-up, \textit{Data.Europa.eu} has published the report \say{Report on Data Homogenisation for High-value Datasets} that proposes a method for HVD identification and homogenization \citep{autoHVD}. 
A method for HVD identification and homogenization consists of three steps, namely: (1) dataset identification; (2) identification or development of common data models, controlled vocabularies, ontologies, and APIs to promote interoperability; (3) application of the common data models, controlled vocabularies, ontologies, and APIs to existing datasets. 
Thus, the very first step concerns the identification of HVDs among the data already published in the \textit{data.europa.eu} portal, in the national or local open data portals of the Member States, or in the data portals of their government agencies, such as statistical offices and geographical institutes, which is considered to be a challenging and error-prone task.
To this end the Data.Europe team decided to work with the current status quo of metadata provision by Member States and \textit{data.europa.eu}, proposing a keyword-based protocol for the identification of datasets that can be partially automated. 
The report acknowledges that there are some initial agreements on how to annotate or provide metadata (e.g., by extending metadata models such as the DCAT) for those datasets, so that they are easier to identify, where in the context of the HVD implementing regulation, the \textit{Semantic Interoperability Centre Europe} (SEMIC) team, in collaboration with the Directorate General for Communications Networks, Content and Technology, has prepared usage guidelines on how to use the DCAT-AP for HVDs. This document provides guidelines on how to use the DCAT-AP for a dataset that is subject to the requirements imposed by the implementing regulation. However, these agreements are not yet implemented. The proposed ad-hoc approach/method can serve as a guide for EU Member States in discovering HVD on their portals, however, due to the complexity of the task it is partly semi-automated with a predominance of activities surrounding the process being manual in nature. This makes this process resource-consuming. Moreover, considering the objective of the report, which is to contribute to the OD Directive, it covers datasets and respective categories indicated as HVD by the EC but not individual Member States. 

Overall, the opening of HVDs is an ongoing initiative, within and outside the EU. However, approaches to identifying and assessing HVDs vary significantly, leading to diverse implementation strategies. Therefore, this study focuses on identifying HVDs available on OGD portals for individual portals and nations, based on actual user interest demonstrated through data usage statistics. We aim to develop an automated approach that requires minimal human intervention.

\subsection{Related works}

Let's us now discuss briefly \textit{What approaches are used to determine HVD} as the literature suggests.

Generally, scientific literature suggests that approaches used in the research can be classified into five groups: (1) pre-defined list of datasets at (1.1) regional level \citep{Shadbolt,varytimou2015towards,Nikiforova2021,tseng2024unlocking} and (1.2) national level \citep{Utamachant,tseng2024unlocking}; 
(2) global initiatives or frameworks (Government Open Data Index (GODI) and Open Data Barometer (ODB), the Open Data Inventory (ODIN), G8 Open Data Charter) \citep{Utamachant,tseng2024unlocking};
(3) consultations with actual users \citep{zsarnoczay2023community,Nikiforova2021}; 
(4)  compliance with a certain characteristic (e.g. real-time data; geospatial data; LOD; business data) \citep{varytimou2015towards,tseng2024unlocking}; (5) mixed approach  \citep{Utamachant,Nikiforova2021,tseng2024unlocking}.

The approach towards HDV determination also depends on the OGD lifecycle phase, thereby being ex-ante or ex-post. 
The former category is less popular in research, mainly due to the complexity of the issue, where it is difficult to predict the potential usefulness of datasets before their opening, and the fact that once identified, the corresponding data owner is not yet identified. 
In the lack of understanding on how to tackle this issue, research-oriented towards this direction predominantly utilizes consultations as the main method \citep{zsarnoczay2023community,Nikiforova23} with ongoing discussions on the need for having a more robust and preferably (semi-)automated approach to be used instead \citep{huyer2020analytical,Nikiforova23,prieto2019framework}. 

Ex-post HVD determination, in turn, is more popular and uses a greater variety of methods. 
In addition to methods such as mapping datasets available on the OGD portals to categories considered to be HVD (in line with the EU OD Directive HVDs \citep{Nikiforova2021}, the global initiatives or frameworks such as the Global Open Data Index (GODI), the Open Data Barometer (ODB), the Open Data Inventory (ODIN), or G8 Open Data Charter \citep{stuermer2016measuring,Shadbolt}, Stuermer et al. uses a more complicated approach by employing a Social Return On Investment (SROI) framework to evaluate the impact of already published datasets \citep{stuermer2016measuring}. This approach can be compared to an “\textit{Impact Assessment study on the list of High-Value Datasets to be made available by the Member States under the Open Data Directive”} \citep{european2020impact} suggested by the European Commission, which considers both economic benefits, environmental benefits, generation of innovative services and innovation (innovation and artificial intelligence), reuse, improving, strengthening, and supporting public authorities in carrying out their mission, resulting in 126 possible indicators classified in 32 categories to measure this value, some of which, however, are qualitative, making this evaluation even more complicated. Moreover, most of these indicators require the collection of supporting data, the volume and complexity of which may be greater than the original dataset  \citep{Nikiforova23}, e.g., \textit{"Purchasing power of consumers," "CO2 emissions," "Quality of healthcare services}" etc.).
These studies mostly identify what individual datasets of those already available can be denoted as HVD.

While none of the studies has reported consultations to be used as a method at the national level, a more in-depth analysis of HVD determination and opening process within countries reveals that more methods, including consultations may be used at lower levels such as local government and individual public agencies responsible for HVD determination among their data (that would correspond to that defined at the national level) \citep{tseng2024unlocking}.

Approaches and determinants used by the research can be categorized into qualitative and quantitative. Some indicators or determinants can quantify the value or potential interest in a dataset, while others are more qualitative, such as citizens' awareness of an environmental issue which can be quantified to some extent but remains fundamentally qualitative (also in line with systematic literature review findings of \citep{Nikiforova23}).
 Depending on the source of these determinants, they can also be divided into \textit{internal} and \textit{external} indicators. Internal indicators are derived from data that the data publisher or the national open government data portal owner possesses, such as usage statistics - views, downloads, number of showcases/use-cases per dataset as part of metadata, usage statistics as part of Google Analytics or portal log files, where the presence and amount of indicators vary per portal depending on its design. External determinants involve input from external actors or stakeholders  - both individuals and systems, such as the G8 Open Data Charter,  Open Data Directive, indices and benchmarks (GODI, ODIN, ODB, ODMR) used to understand potential HVD or to inform decision-making regarding such datasets.

This also aligns with \citep{bendechache2023systematic}, who conducted a systematic review of research exploring the concept of data value and its application in decision-making processes. They found that despite its conceptual origins dating back to at least 1980, the field remains immature, lacking commonly agreed terminologies, models, and approaches. This deficiency inhibits the generalization and common understanding of data value dimensions, hindering further quantification. In addition, only half of the reviewed studies provide data value metrics. Among those that do, five key dimensions emerge: namely economic (also financial), content (also uniqueness), quality, usage, and utility.

Finally, according to \citep{Shadbolt}, HVD identification is only possible when their demand side is understood. To this end, authors stress the importance of data interoperability, which is expected to contribute to the identification of HVD (however, a clear idea is not provided on how exactly this will be then done). In our case, although deviating from the original idea of the authors, acknowledging the importance of understanding the actual demand and/or interest in the existing data, which can then fuel HVD determination, we seek to homogenize the categorization of HVD among different portals within the country (or region) thereby seeking for a more comprehensive understanding of actual HVD state within the area.

\section{Material \& Methods}
In this section, we define the research questions, briefly describing the methods we will use to answer them with more detailed elaboration in the subsequent section. We also define sample and data collection methods.

\subsection{Methods}

Considering the objective of this study, the initial and niche state in which research on HVD is positioned, and the scarcity of methodological results and established shared practices as discussed in the previous section, we define four research questions (RQ). 

(RQ1) \ReQ{What levels of interest do users show in OGD portals?}

Recognizing that understanding the actual demand and interest for OGD data is a fundamental component for the identification of HVD according to the ex-post approach we are proposing, this research issue seeks to quantify this interest through usage indicators such as the number of views and downloads. 

To answer this question efficiently, given the large number of datasets published by different government organizations and countries, it is essential to have an automated procedure that can extract usage information from each published dataset. To this end, using the metadata discovery APIs provided by OGD portal platforms, we developed an application in Python to retrieve portal usage data (i.e. number of views and downloads). The metadata content is then extracted and stored for later analysis.

(RQ2) \ReQ{How can thematic categorization and impact assessment of OGD be conducted to optimize its value and relevance for regions or countries?} 

This research question aims to assess the prominence of specific categories/ topics of OGD in comparison to others within a selected region. 

To answer this question, we first, examine \textit{how thematic information associated with OGD can be automatically extracted}, with further examination of \textit{metrics that can be used to measure the impact of each theme for a selected region/country}. 
Within this research question, we analyze and compare metrics derivable from the retrieved HVD indicators to evaluate the relevance of a given category. As part of this RQ, we define the High-Value Data index (HVDi) supposed to give significance to the absolute values of the published datasets. To this end, we develop code for both extracting thematic data from the metadata and implementing various metrics, which are automatically visualized to improve the perception of results.

(RQ3) \ReQ{How differences and similarities can be identified among High-Value Data across various regions/countries?}

Recognizing the diversity of categorization choices between portals, this research question seeks to 
standardize the categorization of HVDs between portals within a country (or region).
 This standardization aims to facilitate meaningful comparisons despite variations in denotations, thereby promoting a more comprehensive understanding of the actual status of HVDs within the area.

To address this question, two interrelated tasks must be performed:
(i) identifying the categories that are most prevalent across a set of portals, and (ii) aligning the categories of individual portals with those that are most prevalent overall within the studied area, e.g., country.
 To address these challenges, we reuse the methodology proposed by Pinto et al. \citep{Pinto20}, which provides an open-source solution that has been thoroughly tested and could be seamlessly integrated into our existing code.

(RQ4) \ReQ{What are the high-value datasets of U.S. municipalities?}

This RQ aims to identify the HVD of U.S. municipalities as a use-case selected for this study based on the prevalence of their utilization within the sampled municipalities. This RQ allows us to verify the proposed framework, as well as to demonstrate it in action. It can be seen as a byproduct of answering the above RQs by validating the proposed framework on the real-world example. 

As such, we employ an exploratory case study approach \citep{yin2018case}, which is found useful when the issue of study is a contemporary issue with limited empirical information available, and there is a need to develop new propositions \citep{yin2018case,chopard2021methods,eisenhardt1989building}.

\subsection{Sample and data collection}

To evaluate the effectiveness of our methodology, we analyze nine (9) U.S. city portals, all operating on the Socrata engine, serving as our study sample. There are several reasons for this choice.
 First, the United States has been at the forefront of public data openness initiatives since 2009. As a result (and in line with \citep{merrilldouglas2010}), many U.S. municipal governments became early adopters of Open Data practices, spurred by President Obama's administration. As a consequence of this movement and the substantial efforts invested by U.S. cities in publishing Open Data and encouraging citizen usage, numerous scholars have investigated the characteristics of these portals \citep{Zencey2017,MERGEL2018622,pinto18} and scrutinized their performance across various parameters \citep{Barbosa2014StructuredOU,zhu2019an,THORSBY201753,quarati2023}.
It is thus imperative to evaluate potential variations among administrations in cataloging their datasets and the possible impact of these practices on citizens' utilization of them.

Additionally, the underlying assumption is that sharing a common substrate within the same nation-state, characterized by factors like language, national political-administrative structures, and social discourse, should theoretically foster uniformity in the propensity toward OGD (as also suggested by \citep{huyer2020analytical,Utamachant,Nikiforova23}). Any differences would be due to local factors, which would make them easier to understand. This contrasts with the scenario when comparing approaches and trends toward HVD in different countries, where such basic commonalities may not persist.

Moreover, the decision to opt for portals powered by the Socrata engine was influenced by certain limitations associated with the CKAN engine. While CKAN is widely adopted by many countries worldwide, it doesn't consistently include usage data in its metadata accessible via API. Even when such data is present, it often only relates to the overall number of views \citep{quarati2023}. 
The absence of download data makes relying solely on 'Views' less compelling as an indicator for HVD, especially when gauging user interest as "view". 

Regarding the selection of the nine portals, we considered four parameters: a) city population, b) number of datasets published, c) heterogeneity of the city sample, and d) portal powered by Socrata. We selected the first seven Socrata portals with the largest populations and the highest number of datasets (New York, Los Angeles, Chicago, Dallas, Austin, San Francisco, Seattle) and added Buffalo, which has the third largest number of datasets among US cities.
Finally, we included Honolulu due to its unique location as an island city. As shown in Table \ref{tab:US-portals}, at the moment of data retrieval (May 9$^{th}$ 2024) the nine metropolitan areas, with their 17,000 datasets, cover more than 50 percent of the datasets published by the top 100 city portals in terms of population (based on the 2022 census).

\begin{table}[htbp]
\centering
\caption{US portals included to evaluate the methodological approach on HVD determination, ordered by number of published datasets. Population data from US census\protect\footnotemark.}
\label{tab:US-portals}
\resizebox{0.8\columnwidth}{!}{%
\begin{tabular}{@{}lrlrr@{}}
\toprule
\textbf{City}    & \textbf{Population}        & \textbf{Portal}                         & \textbf{Datasets} & \textbf{Categories} \\ \midrule
Austin          &  974,447    &  \url{https://data.austintexas.gov}   & 4,419     & 22         \\
New   York      &  8,335,897    &  \url{https://data.cityofnewyork.us}  & 3,259     & 11         \\
Buffalo         &  276,486    &  \url{https://data.buffalony.gov}     & 2,365     & 14         \\
Chicago         &  2,665,039    &  \url{https://data.cityofchicago.org} & 1,789     & 16         \\
Los   Angeles   &  3,822,238    &  \url{https://data.lacity.org}        & 1,703     & 26         \\
Dallas          &  1,299,544    &  \url{https://www.dallasopendata.com} & 1,173     & 7          \\
San   Francisco &  808,437    &  \url{https://data.sfgov.org}         & 1,133     & 12         \\
Seattle         &  749,256    &  \url{https://data.seattle.gov}       & 1,094     & 9         \\
Honolulu        &  343,421    &  \url{https://data.honolulu.gov}      & 373      & 7          \\
 \bottomrule
\end{tabular}%
}
\end{table}
\footnotetext{\url{https://www.census.gov/data/tables/time-series/demo/popest/2020s-total-cities-and-towns.html\#v2022}}

The dataset counts range from a few hundred open datasets for Honolulu to several thousand in larger metropolises such as New York (3,259), Austin (4,419), and Buffalo (2,365). 
Notably, the dataset volumes in cities like Buffalo are comparable to or even exceed those found on national portals in various countries, including but not limited to Argentina, Slovenia, Romania, Uruguay, Singapore, and others. 

In terms of thematic dataset categories, we observe significant variability, ranging from 7 categories in cities like Dallas and Honolulu to 26 categories in Los Angeles. This variability highlights a notable absence of coordination among different municipal administrations in establishing a standardized classification policy, e.g., as done by the European Commission as a result of the revision of the European DCAT Application Profile (DCAT-AP) - a specification for metadata records, enhancing semantic interoperability across European data portals - defined 13 data categories that are generally followed by EU Member States, with further extension of the DCAT-AP\footnote{\url{https://datos.gob.es/sites/default/files/doc/file/report_dcat-ap_and_its_extensions.pdf}} application profile to meet their own needs (Belgium, Germany, Ireland, Italy, the Netherlands, Norway, Sweden, Switzerland and Spain). However, the quantitative data alone suggests a wealth of interpretative nuances tied to each metropolis's unique socio-political and cultural contexts.
 The observed diversity in thematic categorization may signify either a lack of consensus or a reflection of the distinctive characteristics of each city. A more comprehensive understanding of these dynamics can be achieved through qualitative examination, which will be undertaken as part of this study.

\section{Results}

\subsection{RQ1: What levels of interest do users show in OGD portals?}
By leveraging the APIs provided by portal engines, it is possible to retrieve portal usage data by appropriately employing the metadata discovery APIs available across diverse portal platforms. As mentioned earlier, the CKAN API may furnish information related to the number of views in a '\textit{tracking\_summary}' metadata field, which includes two values: total and recent, i.e., views in the last 14 days. However, not all CKAN-based portals return usage metadata in the '\textit{tracking\_summary}' field. 
Unlike CKAN, the Socrata Open Data API (SODA) offers a RESTful interface that not only furnishes metadata inclusive of total view counts but also encompasses total download counts up to the current date. The code
 that implements the methodology presented in this study leverages this API to programmatically gather usage data from the portals of 9 US cities. This data is extracted from the metadata associated with each dataset and stored for subsequent analysis phases. Additionally, the code automatically generates analysis results through graphical and/or tabular outputs, facilitating immediate comprehension.

\begin{figure}
    \centering
    \includegraphics[width=1\linewidth]{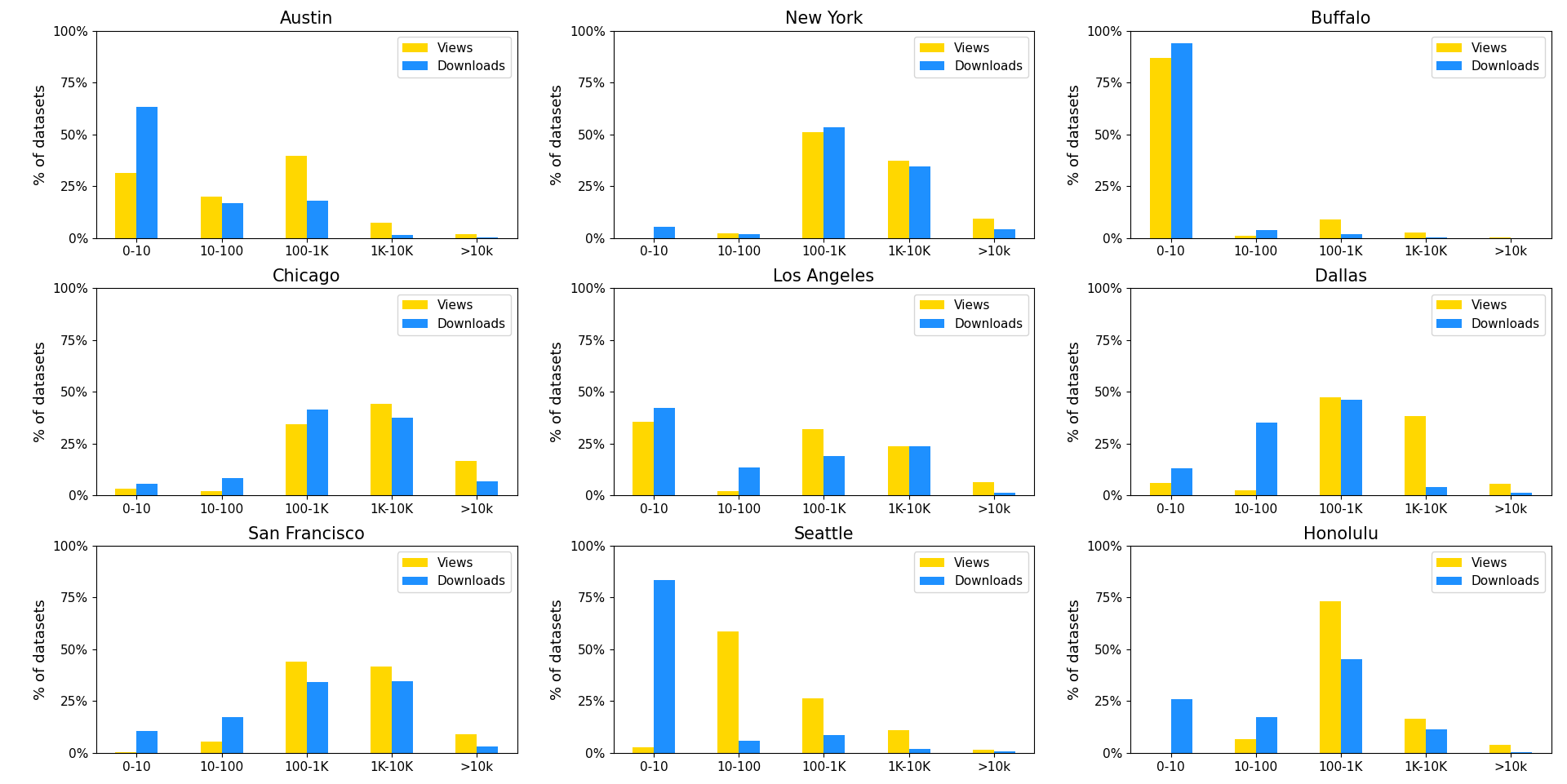} 
    \caption{The overall distributions of Views and Downloads for the 9 US cities. Frequency numbers are grouped into five classes: 0–10, 10–100, 100–1000, 1000–10,000 and >10,000.}\label{fig:ViewsandDownload}
\end{figure}

Figure~\ref{fig:ViewsandDownload} shows the distribution of the number of Views and Downloads for the 9 US portals. 
The X-axis pairs and groups the number of Views and Downloads into five non-linear classes (0–10, 10–100, 100–1000, 1000–10,000, and >10,000), while the Y-axis represents the percentage of datasets within each class.

As we can observe neither Views nor Downloads follow a normal distribution, except for three portals, namely Chicago, San Francisco, and Dallas, which exhibit roughly a normal distribution. Portals of Austin, Buffalo, and Seattle show left-skewed distributions.
 Download descriptive statistics, i.e. min, max, mean, standard deviation, first, second, and third quartile, are provided in Table \ref{tab:downloads-stat}.

\begin{table}[]
\centering
\caption{The overall descriptive statistics for downloads across selected portals 
}
\label{tab:downloads-stat}
\resizebox{0.9\columnwidth}{!}{%
\begin{tabular}{@{}lrrrrrrrrr@{}}
\toprule
\textbf{City}                       & \textbf{Datasets} & \textbf{Downloads} & \textbf{Mean}  & \textbf{Stdd}   & \textbf{Min} & \textbf{1Q}  & \textbf{2Q}  & \textbf{3Q}   & \textbf{Max}      \\ \midrule
Austin        & 4,419    & 5,657,365  & 1,280  & 50,798  & 0   & 0   & 0   & 50    & 2,915,548  \\
New York      & 3,259    & 13,034,572 & 4,000  & 26,419  & 0   & 324 & 696 & 1,888 & 787,782    \\
Buffalo       & 2,365    & 35,021     & 15     & 164     & 0   & 0   & 0   & 0     & 4,695      \\
Chicago       & 1,789    & 25,307,775 & 14,146 & 339,594 & 0   & 180 & 742 & 2,737 & 14,137,087 \\
Los Angeles   & 1,703    & 8,818,120  & 5,178  & 147,920 & 0   & 0   & 55  & 1,000 & 6,065,070  \\
Dallas        & 1,173    & 6,097,138  & 5,198  & 147,786 & 0   & 25  & 107 & 196   & 5,051,324  \\
San Francisco & 1,133    & 6,848,413  & 6,044  & 69,954  & 0   & 66  & 622 & 1,705 & 1,653,082  \\
Seattle       & 1,094    & 445,789    & 407    & 8,011   & 0   & 0   & 0   & 5     & 261,867    \\
Honolulu      & 373      & 189,647    & 508    & 1,140   & 0   & 9   & 151 & 412   & 11,294    
\\ \bottomrule
\end{tabular}%
}
\end{table}

Certain portals stand out for having only one dataset that accounts for a significant proportion of the total downloads. For example, in Austin, approximately 50\% of downloads are attributed to the '\textit{CapMetro Vehicle Positions PB File}' dataset\footnote{\url{https://data.texas.gov/Transportation/CapMetro-Vehicle-Positions-PB-File/eiei-9rpf/}} in the \textit{Transport} category. This contrasts sharply with the 0 downloads in the first and second quartiles and 50 downloads in the third quartile. 
In Chicago, the dataset '\textit{Current Employee Names, Salaries and Position Titles}'\footnote{\url{https://data.cityofchicago.org/Administration-Finance/Current-Employee-Names-Salaries-and-Position-Title/xzkq-xp2w/about_data}} in the \textit{Administration \& Finance} category contributes to more than half (over 56\%) of the total downloads. In addition, two other datasets in the \textit{Service Requests} and \textit{Transportation} categories collectively account for over 3 million total downloads. Chicago's portal also exhibits the highest third quartile (2,737), suggesting substantial interest from local users in OGD.
A particularly striking case of polarisation can be seen in Dallas, where a single dataset - '\textit{Dallas Police Active Calls}'\footnote{\url{https://www.dallasopendata.com/Public-Safety/Dallas-Police-Active-Calls/9fxf-t2tr/about_data}} - in the \textit{Public Safety} category, accounts for over 80\% of downloads from the portal.

Identifying dataset usage trends answers our first research question (RQ1). This step is essential, although not exhaustive, for developing a comprehensive understanding (following an ex-post approach) of HVD determination within a particular region. Subsequently, the next pivotal step involves aggregating datasets based on their categorization, thereby addressing the second research question (RQ2).

\subsection{(RQ2) How can thematic categorization and impact assessment of OGD be conducted to optimize its value and relevance for regions or countries?
}

The organization of datasets into thematic categories emerges as a key feature across OGD portals, aiding user navigation. These categories indicate the thematic scope of the datasets available on the portal and delineate the different domains within which OGD is generated \citep{Pinto20}. This practice is implemented across most platforms. However, not all categories attract equal user attention, as emphasized in the EU report '\textit{Reuse of Open Data}' \citep{EUReport}. Typically, these categories revolve around significant themes such as transportation, economics, government, education, and public safety when describing community activities.

\subsubsection{Extracting Thematic Information}
The metadata retrieved from the portals' APIs allows for the extraction of dataset thematic information based on categories. However, there's a notable discrepancy in the data returned by the CKAN and Socrata APIs. CKAN's metadata includes a '\textit{groups}' field, which may contain multiple associated categories, leading to complexity in categorization and complicating statistical analysis. 
For instance, the national OGD portal of Latvia is powered by CKAN, where examples of datasets such as\footnote{\url{https://data.gov.lv/dati/lv/dataset/groups/2017-gada-republikas-pilsetas-domes-un-novada-domes-velesanu-rezultati-un-veletaju-aktivitate}} can be found, which are classified to multiple thematic areas (\textit{Kategorijas}), which in the case of selected dataset are: \textit{'Valsts pārvalde' ('State administration'), 'Reģioni un pašvaldības' ('Regions and municipalities')}, and \textit{'Iedzīvotāji un sabiedrība' ('Inhabitants and Society')}. This complexity makes it less clear to determine the relative importance of one category over another, thereby impacting the identification of dataset value.

\begin{figure}[!h]
    \centering
    \includegraphics[width=1\linewidth]{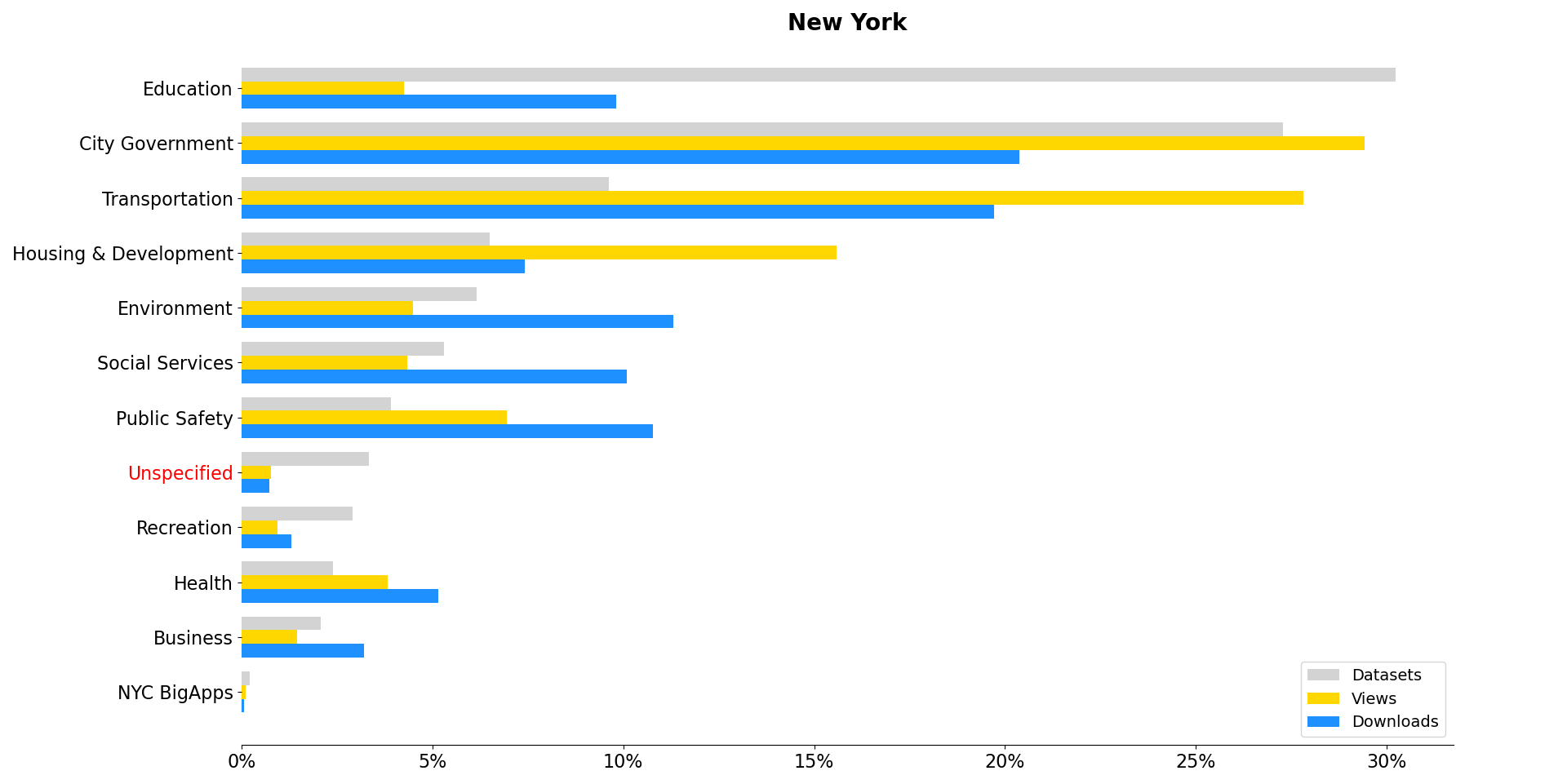} 
    \caption{NY portal categories, ordered by number of datasets per category}\label{fig:NYDatasets}
\end{figure}

In contrast, Socrata assigns each dataset to one category at most, with this information in the '\textit{category}' field, thereby simplifying the categorization process. This streamlined approach enables a more straightforward analysis of the relative significance of categories, thereby enhancing clarity in the identification of HVD topics.
 Figure \ref{fig:NYDatasets} shows the number of datasets, views, and downloads across the 11 categories of the NY portals. It also shows the dummy category '\textit{Unspecified}', encompassing all datasets with unspecified category.

\begin{figure}[!h]
    \centering
    \includegraphics[width=1\linewidth]{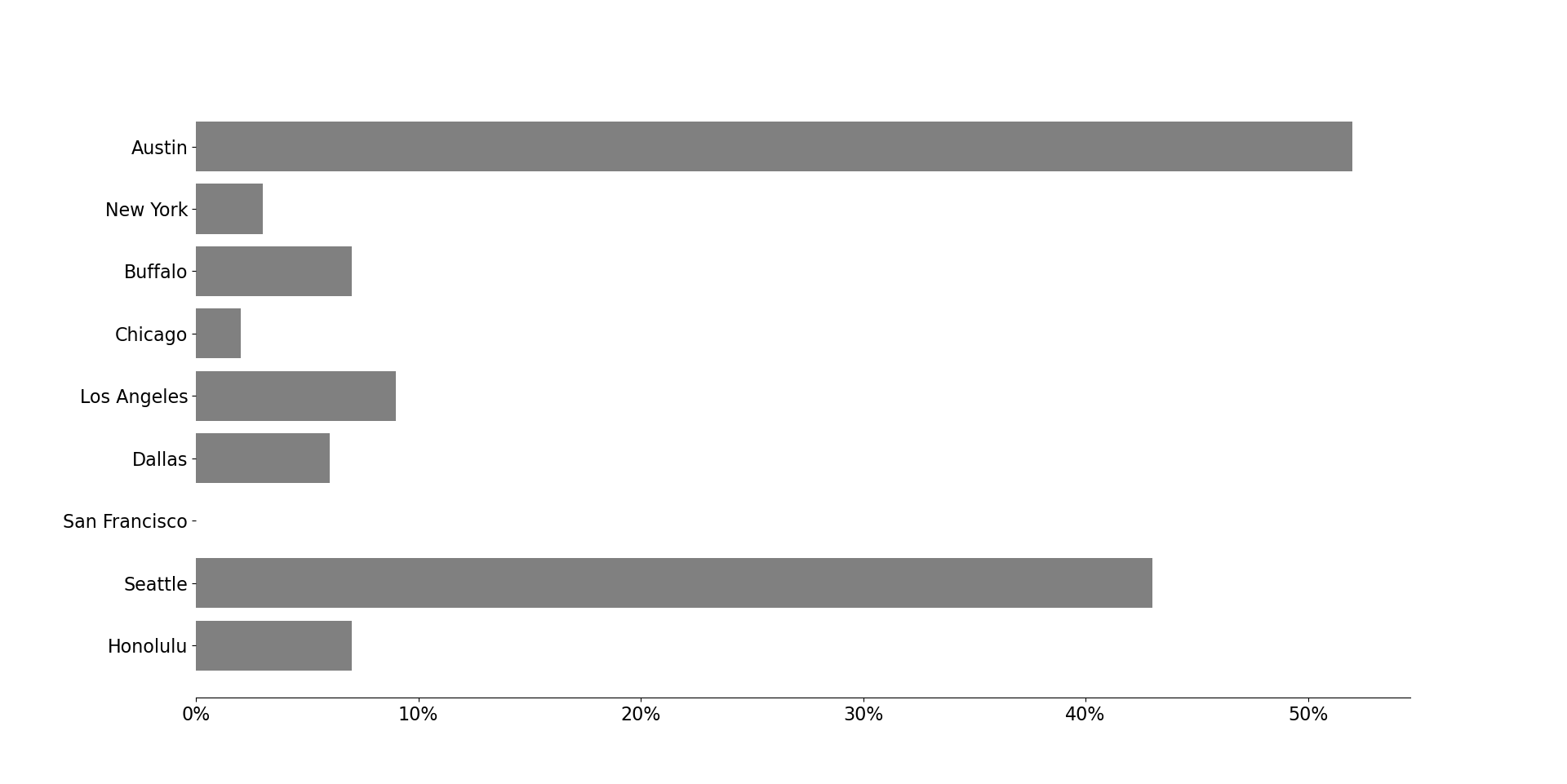} 
    \caption{Percentage of not categorized datasets in the portals sample}\label{fig:Unspecified}
\end{figure}

The situation observed in New York is not an isolated occurrence but rather a common trend. Analysis of the data reveals that a significant portion of datasets across various cities lack categorization, as reported in Figure \ref{fig:Unspecified}. For instance, in Austin and Seattle, around half of the datasets—specifically, 2,310 out of 4,419, and 465 out of 1,094 —remain uncategorized. Less critical is the number of not-categorized datasets for the other cities ranging from 2\% of Chicago to 9\% for Los Angeles, and with  San Francisco with just one not-categorized dataset out of 1,133.
 The situation in Austin is quite surprising, considering the effort made to define 22 categories for organizing datasets that seems to have been overlooked by data publishers, who were expected to allocate their dataset to one of the predefined categories or other actor-person or mechanism - responsible for managing the dataset metadata. This suggests that an excessively detailed level of categorization could sometimes complicate the metadata process. However, a similar situation only occurred to a lesser extent in Los Angeles, with about 10 percent of datasets remaining uncategorized despite the use of 26 categories. Conversely, in Seattle, despite having only 10 categories, 42 percent of datasets remain uncategorized.

\subsubsection{Assessing Thematic Impact}

After extracting thematic information for each dataset, it can be combined with the usage indicator to establish metrics for assessing the potential impact of each category or theme. Various metrics can be derived using simple statistical methods. In the following, we demonstrate this approach using the case of the New York portal. For the sake of clarity, the statistics of the dummy 'Unspecified' category are not reported.

\begin{figure}[htbp]
    \centering
    \includegraphics[width=0.9\linewidth]{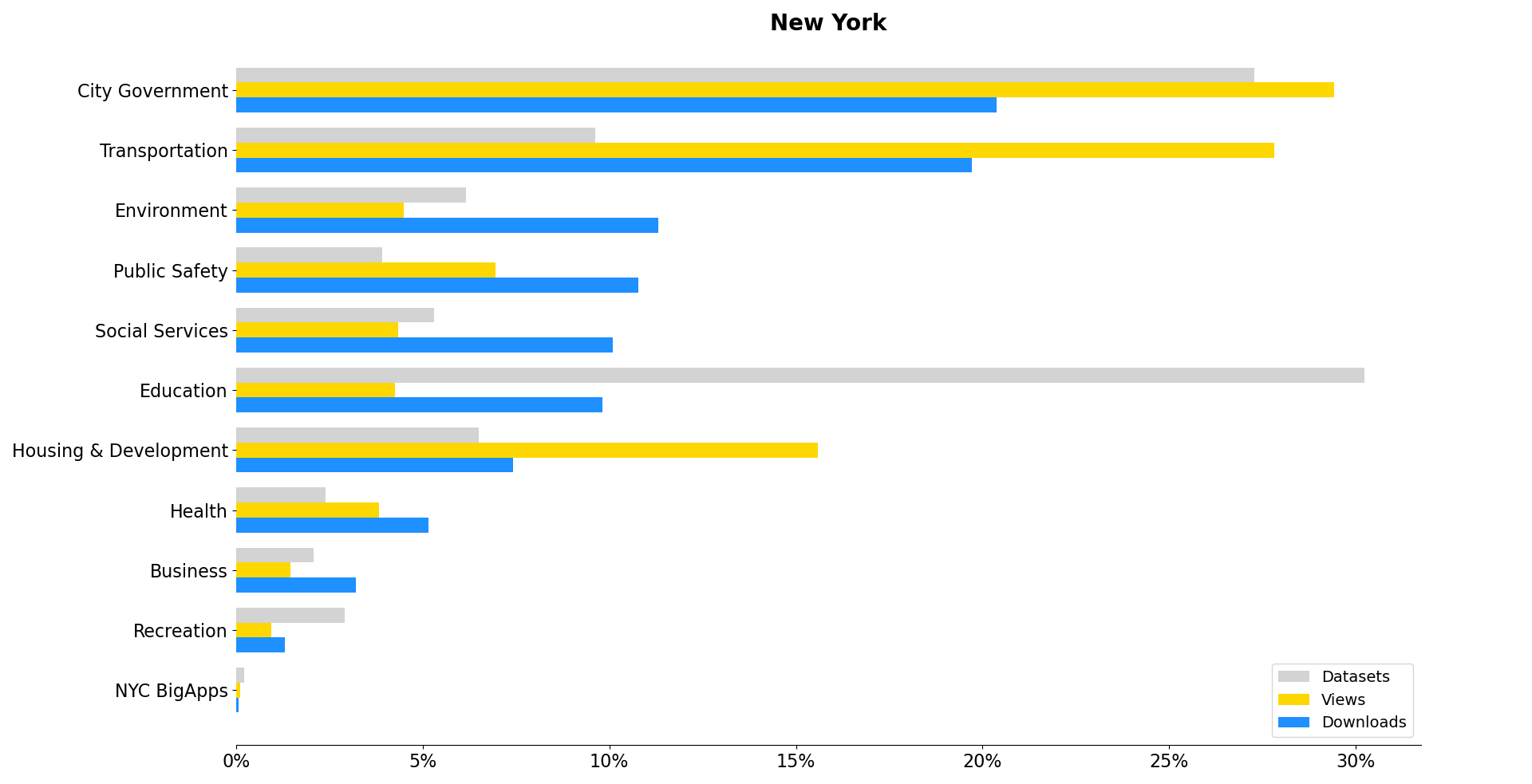} 
    \caption{NY portal categories, ordered by number of  downloads}\label{fig:NYDownl}
\end{figure}

\paragraph{Number of downloads}
A straightforward and rather trivial way is to consider more relevant categories with higher Downloads (or Views) numbers. Accordingly, Figure \ref{fig:NYDownl}, shows that \textit{City government}, \textit{Transportation}, and \textit{Environment} are the categories with higher downloads.

Figure \ref{fig:NYDownl} shows varying levels of interest across categories. Notably, approximately 60\% of total views concentrate on two categories: \textit{Transportation} (28\%) and \textit{City Government} (29\%), which also account for 40\% of downloads. For these two categories, along with \textit{Housing \& Development}, views surpass downloads, as expected. However, in categories such as \textit{Environment}, \textit{Public Safety}, \textit{Social Services},  \textit{Education},  \textit{Health}, and \textit{Business}, downloads significantly exceed views. 
This discrepancy aligns with information from the Socrata support desk\footnote{ (personal communication, November 5, 2019)}, which suggests that automatic tools such as bots or APIs can autonomously download datasets without visiting the portal sites.

Additionally, focusing solely on the Downloads indicator led us to overlook the dataset count for each category. Upon examining results in Figure \ref{fig:NYDownl}, a notable disparity emerges between the number of published datasets and their usage. For instance, \textit{Transportation}, representing only 9\% of published datasets, constitutes about 20\% of downloads. \textit{Environment} with a 6\% datasets attracts over 11\% downloads. \textit{Public Safety}, representing 4\% of total datasets, garners more than double the downloads (over 10\%), and \textit{Social Services} datasets, constituting only 5\%, receive double the number of downloads. In contrast, \textit{Education}, accounting for 30\% of NYC datasets, has only 4\% views and circa 10\% downloads. \textit{City Government}, although with an almost aligned percentage of datasets and views, 27\% and 29\% respectively, has a lower percentage of downloads (20\%) than the published datasets.  
These observations underscore a discrepancy between government data release efforts and proportional user engagement, echoing findings by Berends et al. in \citep{EUReport} (p.26), who identified a \say{mismatch between available datasets and re-used data}.

\begin{figure}[htbp]
    \centering
    \includegraphics[width=0.9\linewidth]{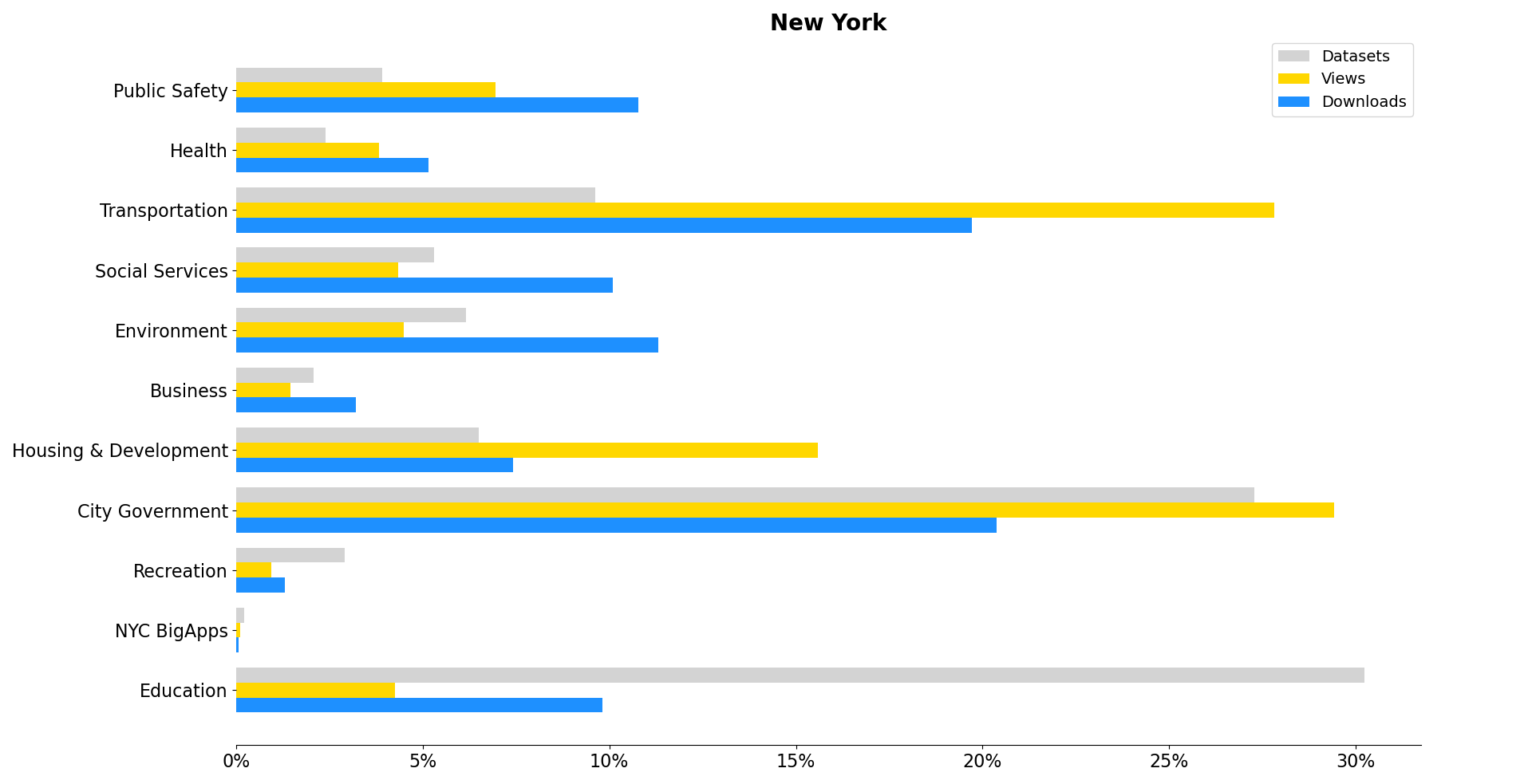} 
    \caption{NY portal categories, ordered by average downloads per datasets (Mean)}\label{fig:NYMean}
\end{figure}

\paragraph{Mean and Median}
To illustrate the interplay between dataset demand and supply for each category, simple statistics such as the mean and median can be employed. Figure \ref{fig:NYMean} shows that, when considering the number of datasets, \textit{Public Safety}, \textit{Health}, and \textit{Transportation} hold the highest rankings, aligning with the previous observations, while \textit{Education} drops from sixth to last place.

\begin{figure}[htbp]
    \centering
    \includegraphics[width=0.9\linewidth]{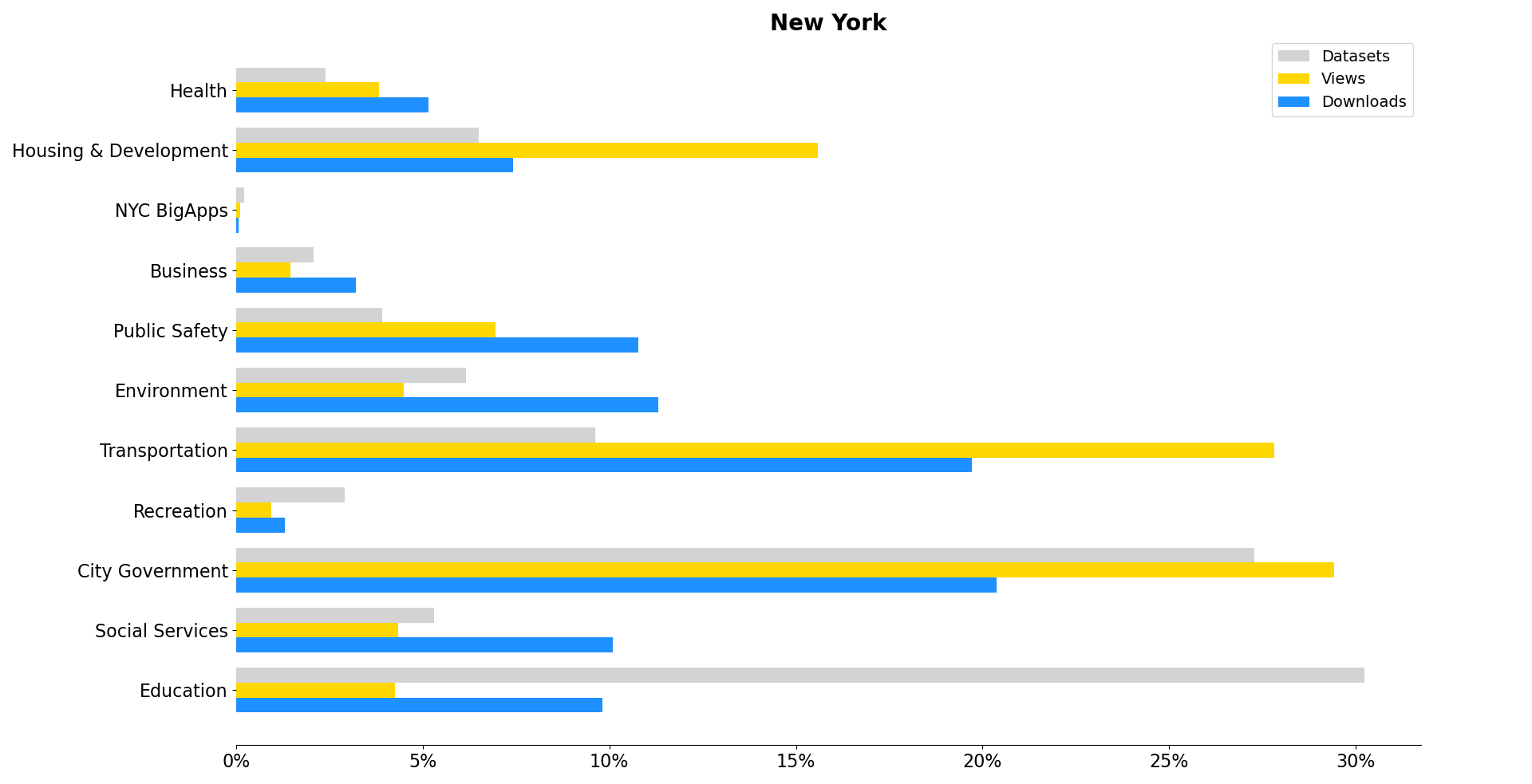} 
    \caption{NY portal categories, ordered by the Median downloads}\label{fig:NYMedian}
\end{figure}

However, the mean of a distribution may not be a robust statistic, particularly in situations with outliers or skewed distributions. This is evident in the usage data depicted in Figure \ref{fig:ViewsandDownload}. In such cases, the median is often considered a more valid alternative.  
As a result, in Figure \ref{fig:NYMedian}, we observe changes in category rankings; for instance, \textit{Public Safety} drops from first to fifth place, and \textit{Transportation} drops from the third to the seventh place, whereas \textit{Housing \& Departments} ascends by five positions.

\paragraph{High-Value Data index}
The median, by mitigating the influence of outliers (datasets downloaded very little or very much, which can heavily impact the average), is often preferable to the mean. However, in our case, it overlooks valuable information crucial for determining the relative 'importance' of one category over another in the eyes of users.

\begin{figure}[htbp]
    \centering
    \includegraphics[width=0.9\linewidth]{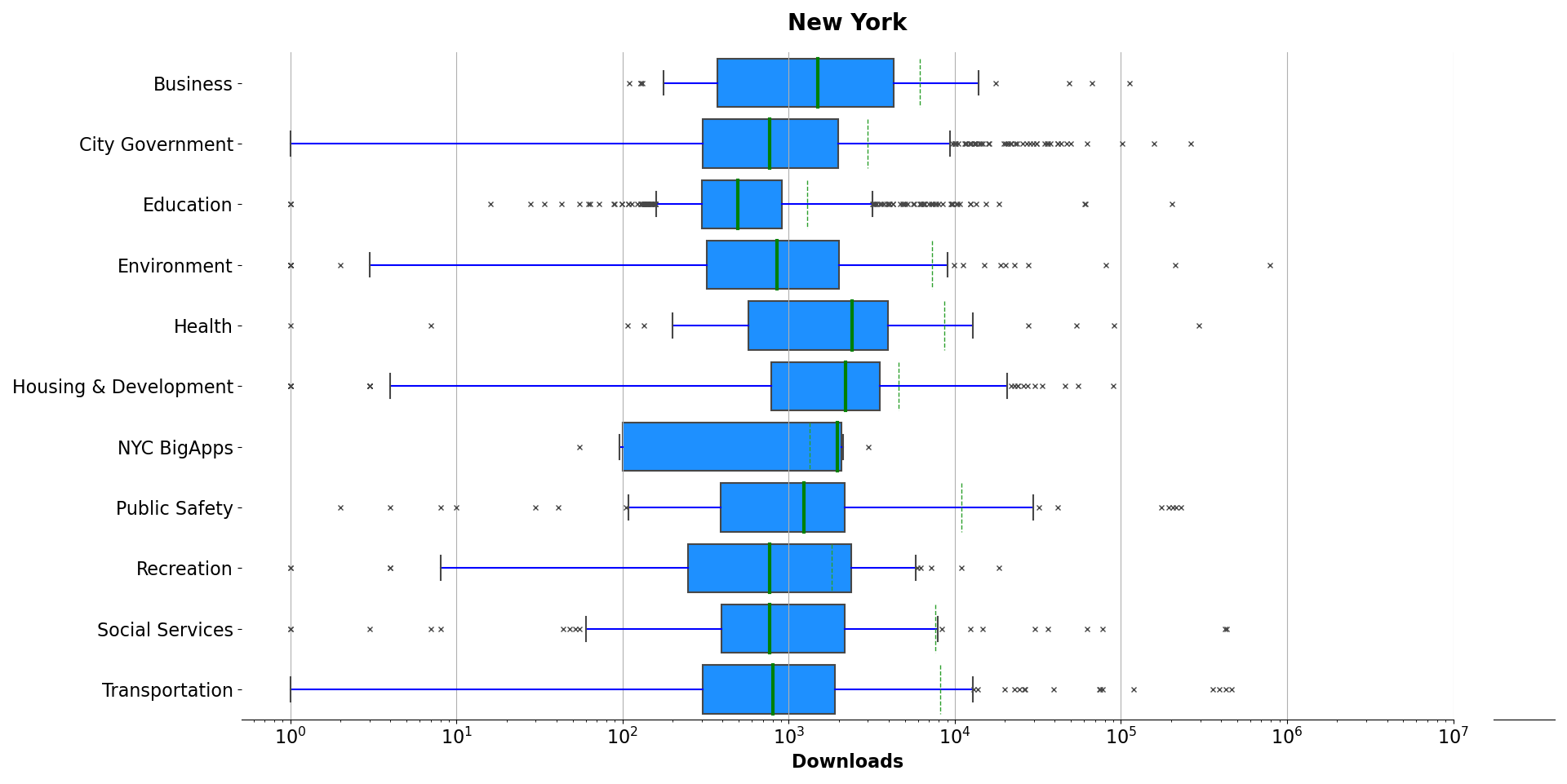} 
    \caption{NY portal categories, boxplot with download statistics. The median is represented by the center line, while the box's limits indicate the 25th and 75th percentiles. Whiskers extend to the 5th and 95th percentiles, and outliers are denoted by crosses.}\label{fig:Boxplot}
\end{figure}

The boxplots in Figure \ref{fig:Boxplot} indicate that while some categories, such as \textit{Education} (which ranks last according to median order, see Figure \ref{fig:NYMedian}), have more datasets published than downloaded, there are still many datasets (the "outliers") that attract public interest, as indicated by their placement to the left of the 95th percentile whiskers. Similarly, the case of \textit{City Government}, with download percentages about two-thirds of the number of published datasets, but with many widely downloaded datasets. It seems reasonable to consider these datasets, which are not few and have received a significant number of downloads, as potentially highly valued. Contrasting this with \textit{Health}, which holds the top position in the median ranking, reveals that \textit{Health} has both a limited number of total datasets and a few datasets that attract substantial downloads.

These observations prompt us to design a metric that retains the robustness of the median while avoiding excessive penalization of categories with numerous widely downloaded datasets, also accounting for the absolute values of the published datasets. Building on this, we define the \textit{High-Value Data index} ($HVDi$) as follows:

\begin{equation}\label{eqHVDi}
{HVDi} = median * \%datasets + \textit{95percentile} * (\%datasets * (1-0.95))
\end{equation}
where \textit{\%datasets} is the percentage of datasets belonging to a category out of the total number of datasets published by a portal, and \textit{median} and \textit{95percentile} are, respectively, the median and 95th percentile values of downloads for that category.

\begin{figure}[htbp]
    \centering
    \includegraphics[width=0.9\linewidth]{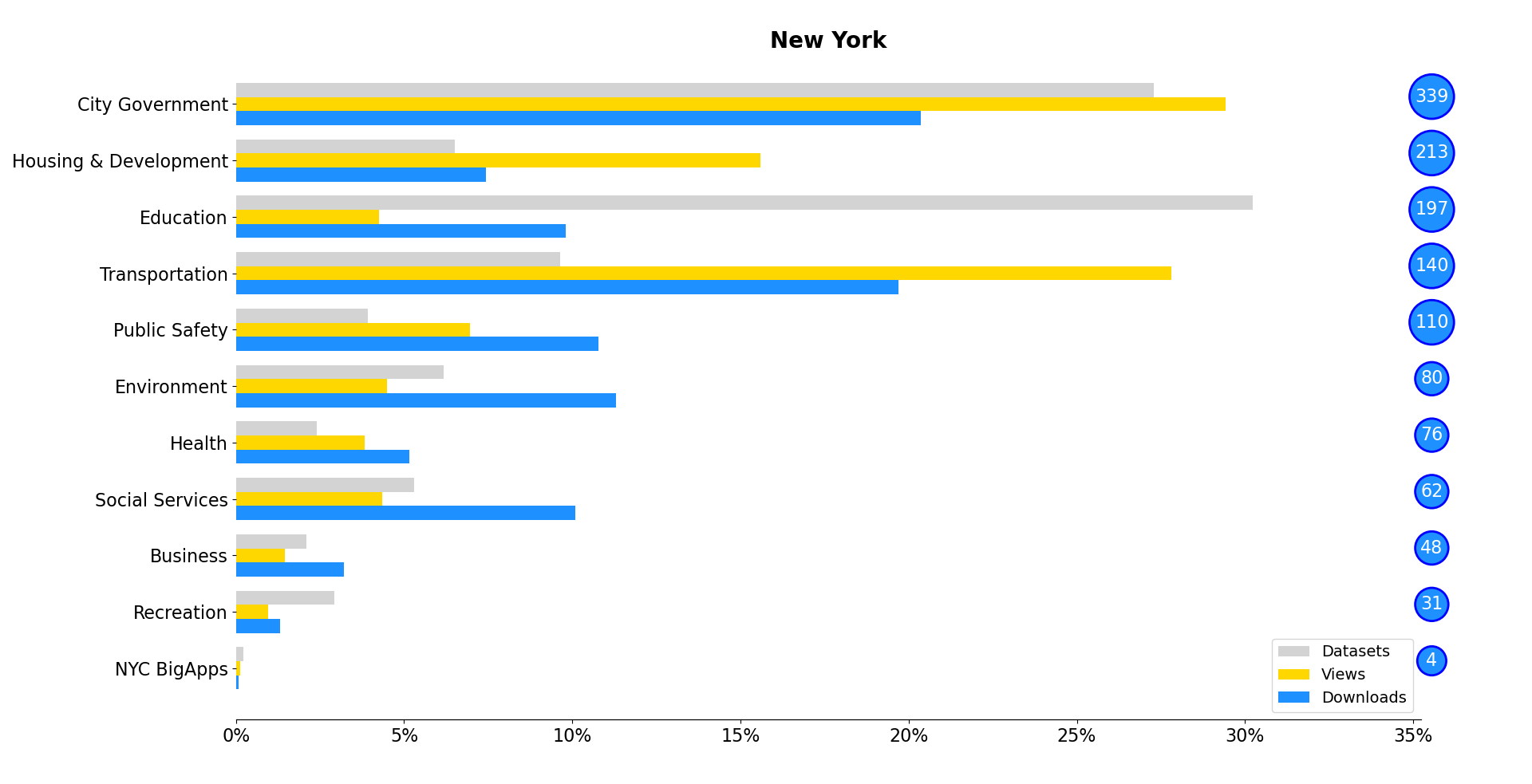} 
    \caption{NY portal categories, ordered by the High-Value Data index, whose value is reported in the right-side bubbles}\label{fig:/NYHDV.}
\end{figure}

The ranking of categories in Figure \ref{fig:/NYHDV.} corresponds to the contribution of the various components in the equation (\ref{eqHVDi}). The top three positions are occupied by \textit{City Government}, \textit{Housing \& Development} and \textit{Education}, all with a substantial number of 'outliers', somewhat mirroring the ranking of these categories obtained from the other metrics examined. 
In particular, \textit{City Government} excels for the number of downloads, \textit{Education} for the number of datasets, while \textit{Housing \& Development}, also ranks second for the median metric, boasting a median value (2,210) that is three to five times higher than that of \textit{City Government} (766) and \textit{Education} (492), respectively. 

The values of the index should not be taken in absolute terms: it probably doesn't make much sense to quantify the value of datasets in the \textit{City Government} category ($HVDi$=339) as three times higher than those in \textit{Public Safety} (110) and four times higher than those in \textit{Environment} (80). From the perspective of a decision-maker (e.g., a portal manager, or a public administrator), evaluating how to organize and potentially reshape the publication of OGD involves a set of criteria that are difficult to summarize in a single number. Much of this evaluation relies on direct knowledge of city issues, citizen information needs, and administrative and managerial policies specific to each municipality. The availability of metrics such as $HVDi$, and the accompanying statistical values, is, in our view, invaluable in activating and guiding a process of management and stewardship of public data, the basis of which those criteria are intended to inspire, and aimed especially at preventing open data as a source of information from being superficial or misaligned with potential users \citep{Santos-Hermosa2023}, ensuring that efforts made for their publishing are not undermined and that the data is effectively reused \citep{quarati2023}.

However, the quantitative value provided by $HVDi$ is necessary when trying to highlight categorical prevalences at the national or regional level, or common to a given set of portals. In other words, when attempting to answer questions like our RQ4 \say{What are the HVD of US municipalities?}. In these cases, a ranking is not equally consistent as it does not capture the dimensional aspect offered by the metric.

On the other hand, the ranking imposed by $HVDi$ on portal categories highlights, as shown in Table \ref{tab:portals_7cats}, the differences that exist among various portals in terms of prevalent categories (compared to $HVDi$). The table displays, sorted from left to right, for each portal the ranking of the top seven categories. The choice of seven is not arbitrary but is because two out of nine portals (i.e., Dallas and Honolulu) categorize their datasets with no more than seven categories. Two aspects are immediately evident: i) some categories are more frequent than others; ii) similar terminologies are used to define the same theme\footnote{Note: For Honolulu, the presence of the 'double' category \textit{business} and \textit{Business} is not a transcription error, but the result returned by the portal's APIs. The same difference does not appear when accessing the portal directly, where exactly only one category \textit{Business} is listed.}.

\begin{table}[htbp]
\centering
\caption{The first 7 categories according to HDVi ranking}
\label{tab:portals_7cats}
\resizebox{\columnwidth}{!}{%
\begin{tabular}{@{}p{2cm}p{2cm}p{2cm}p{2cm}p{2cm}p{2cm}p{2cm}p{2cm}@{}}
\toprule
                                      & \textbf{I}                             & \textbf{II}                               & \textbf{III}                        & \textbf{IV}                            & \textbf{V}                            & \textbf{VI}               & \textbf{VII}                  \\ \midrule
\textbf{Austin}                       & Utilities and   City Services          & Public Safety                             & Building and   Development          & Health and   Community Services        & City   Infrastructure                 & Budget and   Finance      & Transportation   and Mobility \\
\textbf{New York} & City Government                        & Housing \& Development                    & Education                           & Transportation                         & Public Safety                         & Environment               & Health                        \\\\
\textbf{Buffalo}                      & Economic \&   Neighborhood Development & Government                                & Quality of Life                     & Public Safety                          & Infrastructure                        & Economic   Development    & Education                     \\\\
\textbf{Chicago}                      & FOIA                                   & Transportation                            & Public Safety                       & Environment \& Sustainable Development & Health \& Human Services              & Education                 & Administration \& Finance     \\\\
\textbf{Los Angeles}                  & A Prosperous   City                    & City   Infrastructure \& Service Requests & Community \&   Economic Development & Transportation                         & Administration   \& Finance           & Housing and   Real Estate & Public   Safety               \\\\
\textbf{Dallas}                       & Public Safety                          & Services                                  & Archive                             & Economy                                & Other                                 & City Services             & GIS                           \\\\
\textbf{San Francisco}                & Health and   Social Services           & City Management   and Ethics              & Transportation                      & Housing and   Buildings                & Geographic   Locations and Boundaries & Public Safety             & Economy   and Community       \\\\
\textbf{Seattle}                      & Community                              & City Business                             & Transportation                      & Permitting                             & Public Safety                         & Land Base                 & Finance                       \\\\
\textbf{Honolulu}                     & Finance                                & Transportation                            & business                            & Location                               & Business                              & Public Safety             & Recreation                    \\ \bottomrule
\end{tabular}
}
\end{table}

Concerning the first point, it is particularly striking how \textit{Public Safety} and \textit{Transportation} stand out above the others. The former is consistently ranked among the top seven categories (recall that portals such as Austin and Los Angeles classify over 20 categories) for all portals. In comparison, \textit{Transportation} (with the terminological variant \textit{Transportation and Mobility} for Austin) is present among the top seven categories for 7 out of 9 portals. Other recurrent categories in the top seven, but with lower frequency, include \textit{Finance} and similar terms five times, \textit{Health} and similar terms four times, \textit{Education} and \textit{Infrastructure} three times, and so on.

The use of similar terms to define the same theme, as in the case of \textit{Finance} where we find analogous terms like \textit{Budget and Finance} and \textit{Administration \& Finance}, or \textit{Health} also present with \textit{Health and Community Services}, \textit{Health \& Human Services}, and \textit{Health and Social Services}, first and foremost highlights the difficulty, for a human reader, to easily distinguish the real differences resulting from the different topics of the portals (based on a metric) from those that are purely terminological. It also highlights the need to standardize these terminologies to be able to synthesize them, even from an automatic point of view, and to highlight the prevalence of HVD at the national/regional level (as a reading of local behaviors).

\subsection{(RQ3) How differences and similarities can be identified among High-Value Data across various regions/countries?}
Recognizing the significance of grasping the genuine demand and/or interest in the available data, which can subsequently drive HVD determination, we aim to standardize the categorization of HVD across various portals within a country (or a region). This seeks to foster a more inclusive comprehension of the true state of HVD within the area. This standardization process occurs in two phases. The first phase aims to identify the most prevalent categories across a set of portals. The second phase is aimed at aligning the categories of individual portals with those that are most prevalent overall.

\subsubsection{Identifying Most Prevalent Categories Across Multiple Portals}
\begin{figure}[htbp]
    \centering
    \includegraphics[width=1\linewidth]{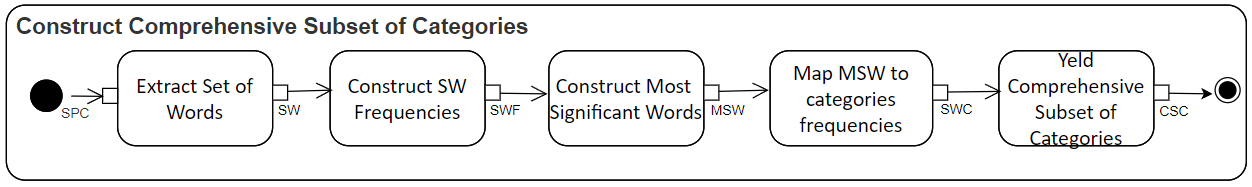} 
    \caption{UML activity diagram for constructing the Comprehensive Subset of Categories given a set of OGD portals (SPC).}\label{fig:/UML_CSC}
\end{figure}

To address the challenges posed by different categorizations in different OGD portals, we adopted the approach of Pinto et al. \citep{pinto18} that proposed a method
 for universal categorization system aimed at improving the organization of datasets in public portals.
 In the UML activity diagram in Figure \ref{fig:/UML_CSC} we illustrate the process consisting of a series of five steps, starting from a Set of Portals Categories (SPC), facilitating the construction of what is referred to as the Comprehensive Subset of Categories (CSC). This subset represents the collection \say{containing the most frequent categories} \citep{Pinto20} in SPC.

The initial step involves extracting the Set of Words (SW) from all categories in SPC, considering that categories typically consist of multiple words. Subsequently, these words are tallied to form the Set of Word Frequencies (SWF), associating each word with the count of portals in which it occurs. Based on this information, the Most Significant Words (MSW) set is identified, comprising the minimal set of words required to attain the highest coverage of portals in SPC.
 Every word in the MSW is linked to all categories associated with it, along with the frequency of their appearance in SPC, yielding the Set of Words-to-Categories (SWC).  
Finally, each word from SWC is associated with the category that appears most frequently in the portals among those containing it.
 This process is repeated for each word in SWC to determine the most prevalent category associated with each word in SPC, yielding CSC.

For the calculation of the CSC, Pinto et al. \citep{pinto18} initially selected a set of 100 active or available US open data portals, corresponding to the most populated cities, according to 2016 US Census data. Given the passage of time and the use of usage data from 2024, we updated this set, using the most recent census data, which ranks the most populous US metropolitan areas (with populations over 50,000). This dataset is made publicly available.
The 100 portals analyzed contain just over 34,000 datasets, with a mean of 365, a standard deviation of 690, and a median of 110. The sample of nine portals, all of which are part of the 100 portals and, including some of the portals with the highest number of published datasets, accounted for more than 50\% of the total datasets.

This update resulted in more than a quarter (26) of the cities affected in the dataset. This was mainly due to changes in population demographics, but also partly due to the activation or deactivation of open data portals. For example, some portals that were previously included are no longer active, while others that were previously excluded (probably due to their unavailability), such as Memphis, Milwaukee, and El Paso, now have functioning open data portals. 
In this update, significant disparities were observed not only in the quantity but also in the definition of categories across many portals.
For example, Los Angeles saw an increase from 4 categories in 2016 and respective study \citep{pinto18} to 27 categories (as of May 2024). Additionally, it was observed that in numerous instances, there had been a shift in the technology underpinning the portals, with the adoption of ArcGIS as the new open data engine. This technological transition also resulted in alterations to the categorization of datasets.

Table \ref{tab:CSC} lists the 32 categories in CSC, extracted from 100 portals of U.S. cities (as a result of the code implementation and its execution over the list of portals),
 arranged in descending order of coverage from top to bottom and left to right.

\begin{table}[!h]
\centering
\caption{Comprehensive Subset of Categories extracted from the 100 portals of American cities}
\label{tab:CSC}
\resizebox{0.7\columnwidth}{!}{%
\begin{tabular}{@{}llll@{}}
\toprule
Public Safety        & Health         & Culture      & Neighborhoods    \\ 
Transportation       & Education      & Finance      & Public Works     \\
City Services        & Parks          & Economy      & Social Services  \\
Recreation           & Business       & Property     & Service Requests \\
Boundaries           & Housing        & Zoning       & Budget           \\
Planning             & Community      & Land Records & Facilities       \\
Environment          & Government     & Demographics & Elections        \\
Economic Development & Infrastructure & Utilities    & Permits          \\ \bottomrule
\end{tabular}%
}
\end{table}

Once the CSC is constructed, the different nomenclature of categories in the sample of nine portals (see Table \ref{tab:US-portals}) can be aligned with those in the CSC. This process allows us to assess differences and similarities in the categorization of datasets across portals, thereby facilitating our answer to RQ4: \say{\textit{What are the HVD of US municipalities?}}.

\subsubsection{Aligning Portal Categories}

\begin{figure}[htbp]
    \centering
    \includegraphics[width=1\linewidth]{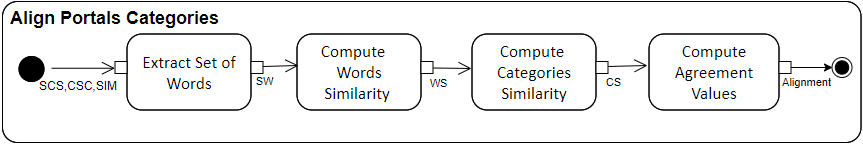} 
    \caption{UML activity diagram for aligning the set of categories in the sample of OGD portals (SCS) with the Comprehensive Subset of Categories }\label{fig:UML_Align}
\end{figure}

To align the categories of individual portals with those that are most prevalent overall, we use the methodology introduced by Pinto et al. \citep{Pinto20}, which we have schematized in the UML diagram in Figure \ref{fig:UML_Align}. The core of this method lies in the establishment of a set of six similarity metrics between each two words, including both edge-counting and content-based methods. These metrics allow us to measure the semantic similarity between two sentences, which in our case represent the categories to be aligned.

The alignment process takes three input parameters: (1) the set of categories in the sample of OGD portals (SCS), (2) the previously calculated Comprehensive Subset of Categories (CSC), and (3) the set of similarity metrics (SIM).
 It is important to note that in this scenario, SCS is strictly contained within CSC. As we described above, CSC was derived from the 100 US portals with the largest population and includes 32 categories (see Table \ref{tab:CSC}). The aim is to encompass the widest and most diverse array of categories used nationally, facilitating the alignment of categories in our sample of portals with a comprehensive representation of categorization practices across US municipal portals.

For each category in SCS, the similarity to the categories in CSC is computed. Initially, the words of the categories to be compared are extracted by tokenization, where each word is separated and stopwords are removed, resulting in a set of SW words. Subsequently, for each pair of words $w_1$ and $w_2$ from two categories $C_1\in$ SCS and $C_2\in$ CSC, the similarity between them is calculated using the six metrics in SIM, yielding WS. The similarity between $C_1$ and $C_2$ is determined based on the maximum similarity among the words in $C_1$ and $C_2$, as produced by each metric in SIM. This generates six similarity metrics for each category $C_1$ and each category $C2\in$ CSC, resulting in a Categories Similarity set (CS). In principle, these values may differ. 
The alignment between $C_1$ and CSC is established by selecting the category in CSC that is agreed upon by the largest number of metrics in SIM. For example, when applying the procedure to the Austin categories, the category \textit{Health and Community Services} is initially associated with the categories \textit{Community} by 3 metrics and \textit{Health} by 3 others. In this scenario, the tie is randomly resolved by aligning \textit{Health and Community Services} with \textit{Community}. Conversely, in the case of \textit{Utilities and City Services}, which is initially associated with \textit{Utilities} (chosen by four metrics) and \textit{City Services} (chosen by 2 metrics in SIM), the first association is selected. 

This process is iterated for all categories of all portals in SCS and concludes by generating the \textit{Alignment} set of pairs (c, c'), where c $\in$ SCS and c' $\in$ CSC, also denoted as $(c \rightarrow c')$ (indicating that c is aligned with c').

\begin{figure}[htbp]
    \centering
    \includegraphics[width=1\linewidth]{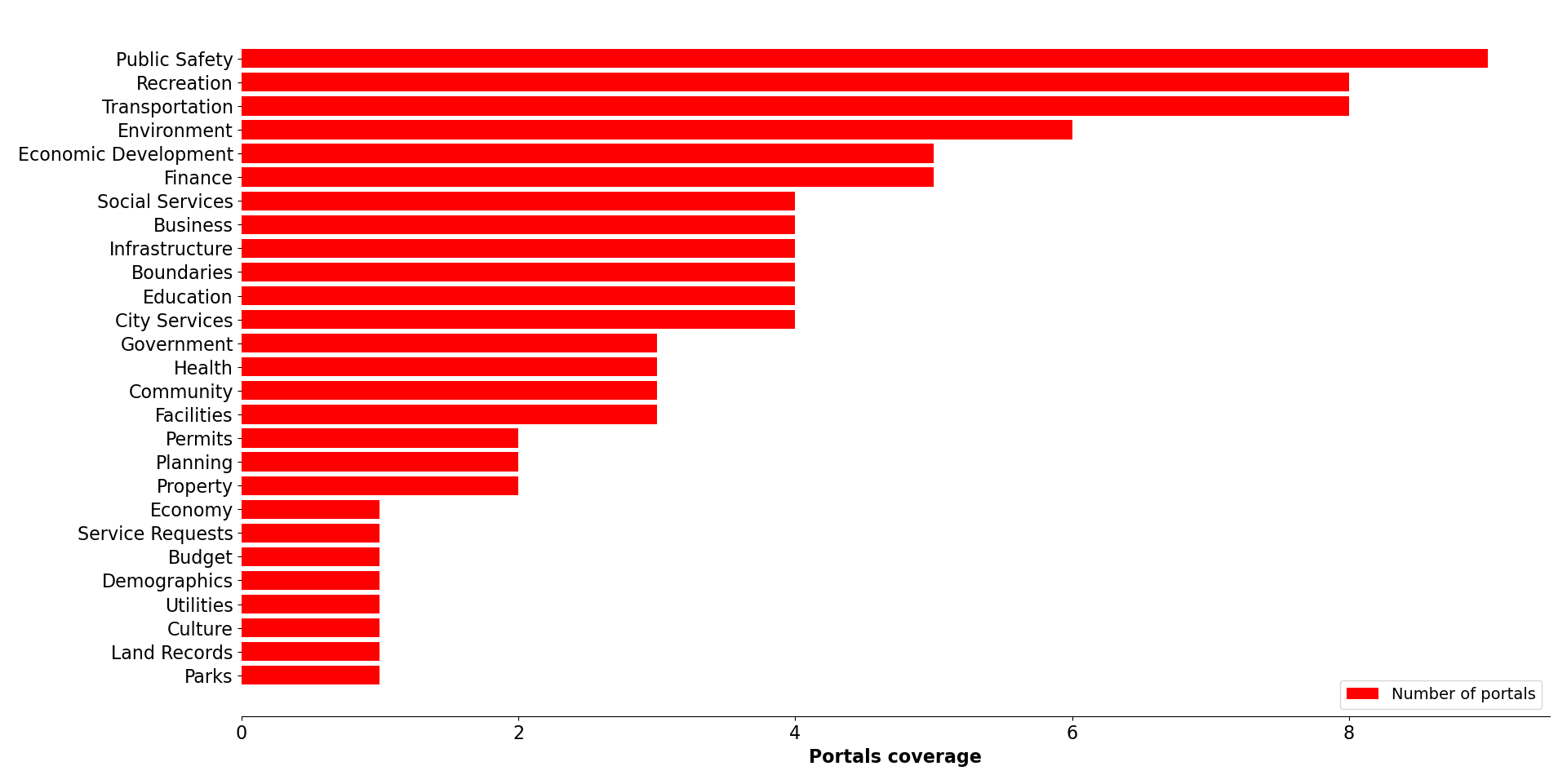} 
    \caption{Sample's portals coverage of the 27 aligned categories}\label{fig:/Cat_Portals_stats}
\end{figure}

The result of the alignment process is a mapping in which 27 out of 32 categories of the CSCs match those of the 9 sample portals. Figure \ref{fig:/Cat_Portals_stats} shows the frequency with which the CSC categories are aligned by the categories of the 9 portals. The graph gives a first overview of the most frequent categories. It is worth noting that the first 6 categories are present in over 50\% of the sample. This observation is unsurprising, considering that, except for \textit{Finance}, the other 5 categories also rank among the top 8 most frequently used categories, as reported in Table \ref{tab:CSC}. 
Additionally, Table \ref{tab:portals_coverage_alignmnet} in Appendix \ref{appendix} illustrates the alignment of the categories in CSC with those of the sample portals.

\subsection{(RQ4) What are the HVD of US municipalities?}

Once the various terminologies used to categorize datasets have been unified into a common framework, it becomes feasible to address RQ4. 
Utilizing the set of categories in the Comprehensive Subset of Categories ($CSC$), we apply equation (\ref{eqHVDic}) to determine, for each category $c \in CSC$, its high-value data index $HVDi_c$ based on the $HVDi_{\tilde c}^p$ indexes computed, by equation (\ref{eqHVDi}), for each portal $p$ in the pool of selected portals $P$, and for each  $ \tilde c : \exists ({\tilde c} \rightarrow c)  \in Alignment$.

\begin{equation}\label{eqHVDic}
    HVDi_c = \sum_{p\in P} HVDi_{\tilde c}^p * |d^p|/|D_P| 
 \end{equation}

where, for each category $c\in C^p$ of a given portal $p \in P$ organized as a set of categories $C^p$, we have $HVDi_c^p$ defined as:

$HVDi_c^p = median_c^p * \%datasets_c^p + 95percentile_c^p * (\%datasets_c^p * (1-0.95))$

where 
$\%datasets_c^p=|datasets_c^p|/|d^p|$, with $d^p$ is the set of datasets of a portal $p$.  And with $D_P$ denoting the overall datasets in the sample $P$.

\begin{figure}[htbp]
    \centering
    \includegraphics[width=1\linewidth]{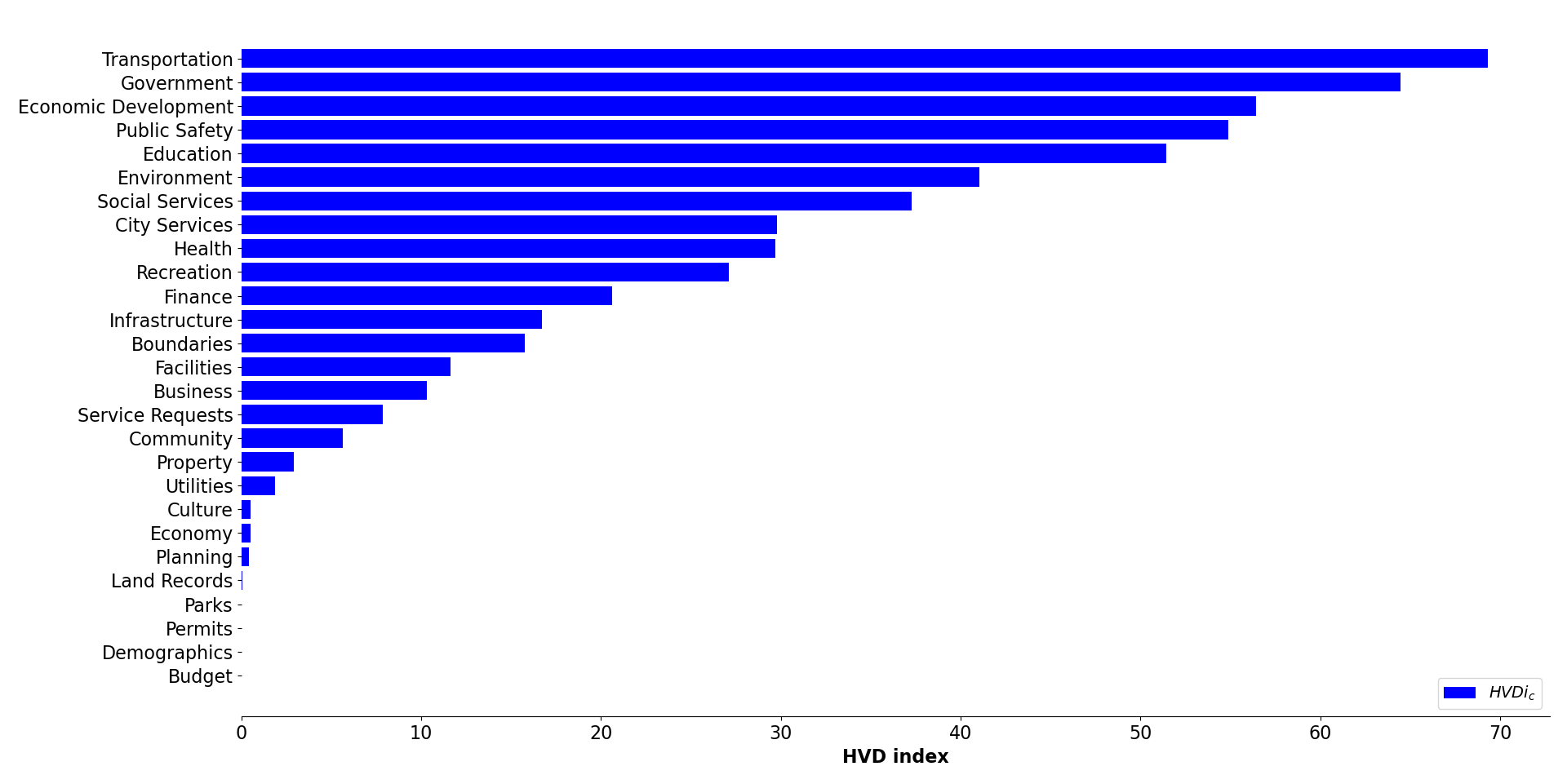} 
    \caption{HVD categories of the 9 portals sample (based on the $HVDi_c$ index)}\label{fig:/HVD_stats}
\end{figure}

The results of these calculations are provided in Figure \ref{fig:/HVD_stats}, from which several observations can be made. Considering the absolute value of $HVDi_c$, we observe that for the first seven categories, this value at most doubles (ranging from 37 for \textit{Social service}, to 69 for \textit{Transportation}), gradually diminishing for the remaining categories, with marginal results (below 10) for 12 categories. A superficial reading shows that the top 15 are practically all categories that one would expect to be in these positions since they are among those most present in city portals, not only the US, and therefore reasonably more sought after by users. The value of $HVDi_c$, reflects this situation considering that it is based on the weighted average of the $HVDi_c^p$ across all portals where the category (or rather an alignment of it) is present, and the latter is closely associated with downloads, both in terms of median and the number of datasets exceeding the 95th percentile.

If we examine the ranking determined by $HVDi_c$ and compare it with the ranking determined by the frequency of each category across the nine portals, as shown in Figure \ref{fig:/Cat_Portals_stats}, we notice that of the top six categories in the latter figure (with a coverage >50\%), four also appear in the top six positions for $HVDi_c$ (i.e., \textit{Transportation}, \textit{Economic Development}, \textit{Public Safety}, \textit{Environment}), although not necessarily in the same order. \textit{Recreation} and \textit{Finance}, which appear in the top six of Figure \ref{fig:/HVD_stats}, are not among the top 6 in Figure \ref{fig:/Cat_Portals_stats} and are replaced by \textit{Government} and \textit{Education}. However, we observe that eleven of these 15 categories in terms of $HVDi_c$ ranking are also in the top 15 for coverage.

These considerations reinforce our opinion that the identification of High-Value Data should not solely be based on how prevalent a category is across the considered portals, but should also take into account appropriately selected usage parameters underlying metrics such as $HVDi_c$, which can somehow capture users' interests.

\section{Discussion, Limitations and Future work}
Compared to the current state of the literature and the directives, guidelines, and efforts of various governmental bodies, in terms of practices aimed at the definition, identification, and implementation of HVDs, our methodological proposal aims to automate the identification of HVDs available on OGD portals through a quantitative approach, based on the actual interest shown by portal users, as gleaned from data usage statistics and seeks to minimize human intervention as much as possible.

\begin{figure}[htbp]
    \centering
    \includegraphics[width=0.5\linewidth]{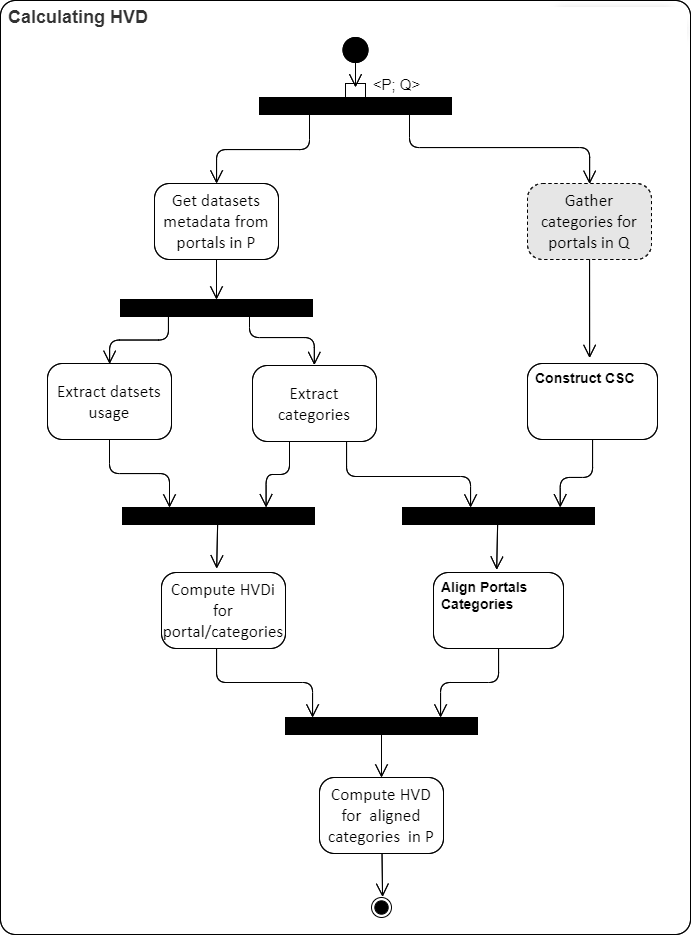} 
    \caption{UML activity diagram illustrating the Usage-Based HVD methodology under the four defined research questions. Regarding our example, set P comprises the 9 US portals, while set Q consists of 100 US portals from which to derive the Comprehensive Subset of Categories. Generally, set P is a subset of set Q. The two UML diagrams of the two sub-activities 'Construct CSC' and 'Align Portals Categories' are shown in figures \ref{fig:/UML_CSC} an \ref{fig:UML_Align}.}\label{fig:/UML_HVD}
\end{figure}

As a demonstration, we applied our methodology, developed in response to the research questions, to a sample of US municipal portals. This allowed us to showcase the different steps involved and the techniques and methods used. According to the workflow depicted in Figure \ref{fig:/UML_HVD}, these steps can be summarized in four main phases: i) downloading the metadata of datasets from selected portals and extracting the usage information of the assigned categories; ii) calculating the metrics (i.e., $HVDi$) for each portal; iii) standardizing and aligning the categories; iv) computing the $HVDi_c$ for the aligned categories. 
To implement the workflow, we integrated existing technologies for the standardization phase with ad hoc solutions developed for steps i), ii), and iv). This makes the entire process automated, except for extracting categories from a larger set of portals (in our case, data related to 100 US portals), as indicated by the dashed gray box, which may require some manual intervention. 
The open-sourcing of the code aims to contribute to the replicability of the method in similar contexts, following the workflow outlined in Figure \ref{fig:/UML_HVD}.

From a practical viewpoint, the proposed methodology can find application in two different areas, and the process outlined in Figure \ref{fig:/UML_HVD} can lead to two types of paths: a 'partial' one that concludes with the action "Compute HVDi for portal/categories", and the other that concludes at the end of the process. In the first case, the use of a metric like $HVDi$ will allow a 'local' decision-maker (for example, a city portal manager, a public administrator), in an ex-post setup, to assess \textit{whether and how to reorganize the categories of published datasets}. 
Consequently, the question of \textit{whether redistribution of the efforts of publishing datasets is required} can be answered, based on the indications derived from their use. As such, $HVDi$ will serve as a stimulus and guide for processes of stewardship of public data and can complement direct knowledge of local issues, citizen information needs, and specific administrative and managerial policies of municipalities. 

The computation of aligned HVD categories, obtained as a result of the entire process, allows a higher-level understanding of differences and similarities in data usage across different regions/countries, facilitating not only a better understanding of the dynamics themselves but also the possibility of implementing shared - or rather ad-hoc - practices within the context of organizing and managing 'local' portals (e.g., municipal, regional portals) under the political and/or managerial supervision of a supervisory governmental entity (e.g., national portal), including seeking for greater homogeneity across portals within an area, e.g., country, and as such increased interoperability. 

From our analysis, we noticed considerable variations among US municipality OGD portals concerning the selection, number, and terminology of categories used for dataset grouping. This resonates with previous studies by Zencey \citep{Zencey2017} and Pinto et al. \citep{pinto18}. The former, focusing on the prevalence of topics among datasets released by 141 US public bodies, noted that \say{popular datasets varied significantly based on location} \citep{Zencey2017}. This divergence, as Zencey suggested, reflects local preferences and demands. The latter conducted a study on how the 100 different portals of the most populous cities in the United States categorize their datasets and noted that users wanting to perform a comparative study between different cities within a state or country could face difficulties navigating the varying categories or topics of each city's portal \citep{pinto18}. It is noteworthy that, unlike our contribution, both these studies do not delve into portal usage data or identify trends based on detailed usage statistics.

We identified, that although most portals have a rich list of predefined categories, there are still datasets that are not assigned to any category. Moreover, while for some portals, the ratio of these datasets to the total number of datasets in the portal is small (from the 2\% of Chicago to 9\% for Los Angeles, and 1 of 1,133 datasets in San Francisco), for some portals such of Austin and Seattle, this applies to more than half of the datasets, even though the former portal has over 22 predefined categories for organizing datasets. This trend is not unique to U.S. municipalities portals, which is also in line with the case of the Latvian national OGD portal, as reported in \citep{nikiforova2020assessment}, where a similar trend was observed. Moreover, since the OGD portal of Latvia is powered by CKAN, identifying datasets that are not assigned to any predefined category is more difficult since CKAN compared to Socrata, by which U.S. portals are powered, does not have a pre-requisite for assignment strictly to one category. CKAN-powered portals, in turn, do not pre-determine the number of categories to be specified for a dataset, which can span from zero to several categories. 
This also reveals another potential area of research for the OGD community, where mechanisms for automated dataset classification and assigning to categories can be implemented thereby augmenting or automating the process of metadata definition for data publishers. In light of current technological trends, Large Language Models (LLM) can find application here. Similarly, this can find applications for Socrata portals, thereby assigning datasets to pre-defined categories or suggesting alternatives, if an inaccurate assignment of categories is revealed.

When comparing the list of HVD topics as identified by the EU and in our study for the US, we can see significant overlaps, but at the same time some unique topics within both regions. Both the EU and the US recognize the critical importance of health-related data, underscoring the universal need for robust health data to inform public policy and improve healthcare outcomes. Additionally, both regions value data related to government and public administration, transportation, and mobility, reflecting the essential role of transportation in economic development and urban planning. 
Unique thematic areas in the EU include geospatial data, meteorological data, statistical data, and a strong emphasis on earth observation and environment data, which involves environmental monitoring and earth observation to support climate action and sustainable development. Furthermore, the EU has proposed additional categories that may be recognized as HVDs in the future. These include climate loss data, which pertains to damages and losses due to climate change, energy data related to production, consumption, and distribution, justice and legal affairs data, and linguistic data, which were not identified among the HVD of U.S. municipalities. However, some unique thematic areas in the US are not explicitly outlined by the EU. These include public safety data, education data, social services data, and city services. These differences, however, may be due to the level both HVD sets consider, whereas the EU focuses primarily on the national level initiatives with some recent increase of interest in local OGD, whereas we study local OGDs of the U.S., which are naturally more concerned about city-related topics. As such, while both the EU and the US recognize the importance of health and governmental data, their priorities reflect regional needs and strategic interests (although alignment of the above-discussed categories as presented in the above section would provide more clarity on the extent of differences). The EU's approach includes a detailed breakdown of environmental and economic data, whereas the US focuses on practical, service-oriented datasets relevant to public services and safety. This underscores different regulatory environments and policy objectives between the two regions, supporting our assumption that identification of HVD should be done at the regional and country level thereby contributing to the local data ecosystem sustainability and resilience.

\subsection{Limitations}
This study has several limitations, some of which, in turn, define the future research agenda. 

First, the developed approach is tailored to OGD portals powered by Socrata, whereas CKAN and ArcGis are also popular data management systems, thereby partly limiting its generalizability. We intentionally decided to focus on Socrata portals as they provide download statistics by default, whereas for CKAN and ArcGis portals it is not the case. As such, even if the portal owners have extended their respective data management systems with download statistics to be part of metadata, the proposed solution will require adaptation to these systems. This is because, in contrast to Socrata metadata, where each dataset is assigned a unique field for category identification, CKAN allows different categories to be associated with the same dataset. This complexity in categorization complicates statistical analysis. Consequently, determining the relative importance of one category over another becomes less straightforward, impacting the clarity in identifying HVD. This means that the proposed solution will need to be extended to tackle the above challenges, probably by employing multilevel modeling for repeated measures. On the other hand, to facilitate this extension, we described the proposed method in detail and made most of its components publicly available (incl. associated datasets of portals).

Second, building upon prior research \citep{Barbosa2014StructuredOU,european2020impact,quarati2023,begany2021understanding} we contend that usage parameters may serve as indicators of interest in a dataset. However, we recognize that mere downloading does not guarantee subsequent dataset reuse \citep{Nikiforova23,quarati2023}, and that true value lies in effective utilization. As such, 'download statistics' serve as a quantitative and '\textit{ex-post}' indicator, contributing to the identification of HVD, which integrated with other methods can ensure a more comprehensive approach to HVD determination.

In addition, download rates, as an indicator, offer retrospective insights (ex-post). While our analysis of OGD portals within a selected area, such as a country, can provide valuable insights into the most valuable categories, suggesting areas where opening more datasets could be beneficial due to greater user interest, it does not prescribe which specific datasets should be opened. In essence, our proposed method primarily focuses on identifying categories that hold the greatest value within the area, thereby streamlining the process of HVD determination. To achieve a more comprehensive understanding of specific HVDs, it would be beneficial to complement our approach with an ex-ante approach. This combined approach would offer a more accurate assessment by incorporating prospective considerations alongside retrospective data analysis. While our method provides valuable insights into category-level trends, integrating an ex-ante approach would enhance the precision of identifying individual datasets with high value, ultimately contributing to more informed decision-making regarding data publication on OGD portals. As such, the proposed method can be perceived as a semi-ex-ante approach, rooted in ex-post assessments. While our methodology primarily relies on analyzing past data usage patterns (ex-post assessment), it also incorporates elements of anticipation and foresight (semi-ex-ante approach) by leveraging these insights to inform future decisions. By examining historical download rates and user interactions with datasets, our approach provides valuable hindsight that can guide proactive decision-making regarding data publication and prioritization. However, to fully embrace an ex-ante approach, incorporating prospective considerations and predictive analytics would enhance the method's ability to anticipate future trends and identify HVD preemptively.

Moreover, the presence or lack of "interest" in datasets expressed in usage statistics - views, downloads, re-uses, can be also attributed to their quality \citep{quarati2023}. In other words, the dataset that is poorly downloaded or used may thematically be compliant with the HVD, however, low usage metrics may be the result of failing to meet expectations content-wise (in-depth or breadth), poorly documented, and low data quality of the content or metadata. 
As such, although this may be recognized as a limitation of the proposed approach, at the same time, our approach can help solve this issue (partly), i.e., when themes found to be compliant with HVD are determined with certain datasets having low downloads statistics, this can serve as an indicator of potential issues with the dataset, triggering their investigation and improvement, should the issue is due to low quality - data quality or metadata quality - poor accompanying documentation, or opening alternative dataset serving a similar purpose.

Although we rely on usage statistics as a quantitative measure thereby seeking for SMART indicator that can contribute to the automated determination of HVD, additional limitations associated with categories that relate to categorical data must be acknowledged. In particular, this relates to the accuracy of the categorization of datasets, which we take for granted in this study. In addition, depending on the accuracy of category definitions within the portal, data cleaning may be found useful. For instance, in the case of the OGD portal of Honolulu, we found the presence of the 'double' category \textit{business} and \textit{Business} as returned by the portal's APIs, which, in turn, does not appear when accessing the portal directly, where exactly only one category \textit{Business} is listed. For the sake of an increased automation of the process and seeking for a streamlined and sufficiently simple process, we have not intervened, however, the risk of the above inaccuracies must be taken into account. At the same time, integrating data cleaning in the process keeping it as automated as it currently is, is another future research direction. 
As we mentioned above, LLM can find applications in this area, where they can be employed to check the accuracy of assigned categories, as well as substitute the steps related to harmonization of categories across portals.

Additionally, due to the selected sample of U.S. municipalities' portals, the results of this study are reflective only of U.S. municipalities OGD that can be considered of high value. 

The dynamic nature of data usage patterns presents a challenge for drawing long-term conclusions regarding HVDs in the United States. As data use patterns evolve, the accuracy of conclusions drawn from this study may diminish, thereby complicating comparisons with future studies, both within the U.S. and in cross-country contexts. Additionally, the inherent instability of OGD portals, characterized by migrations to other platforms or closures, further complicates the research landscape, which we observed when re-used CSC by \citep{pinto18} and \citep{Pinto20}, where we observed instances where dataset links became outdated or led to inactive portals, rendering download statistics inaccurate and compromising overall research outcomes. Addressing these limitations, the need for periodic repetition of this study becomes apparent. However, the developed method and supplementary materials enhance the replicability and reproducibility of this study, facilitating its repetition by both internal and external parties. In addition, the maintenance of associated datasets is crucial to ensure the ongoing relevance and accuracy of research findings amidst portal migrations and closures.

\subsection{Implications}
The study offers a structured methodology for automated identification of HVD across different regions within areas, such as countries, aiding in data reuse and open governance initiatives.

It provides a blueprint for future use by facilitating the identification of datasets with high reuse potential, catering to the unique needs of the selected area, which is the U.S. within this study, and guiding data prioritization efforts.

The study contributes to practical implications by offering insights into disparities and commonalities in dataset utilization, enhancing localized understanding of HVD, and informing open governance initiatives, including seeking an enhanced homogeneity and interoperability of data within a region.

Theoretical contributions include the development of the High-Value Data index (HVDi) to assess dataset categorization across portals, enabling comparisons and identification of HVD within an area (U.S. municipalities in this study).

The study's methodology of automated analyzing of download data from OGD portals to identify high-value categories and compare HVD datasets across different portals within the area provides a novel approach and as such understanding of region-specific HVD, contributes to the theoretical framework valuable insights for data prioritization and open governance initiatives, emphasizing the importance of region-specific insights for effective data prioritization and governance that goes beyond generally acknowledged HVD that oftentimes can be seen as rather "base" OGD.

\section{Conclusions}

Recognizing the importance of understanding genuine demand and interest in available data to drive HVD determination, our study aims to standardize HVD identification across various portals within a country or region. This standardization process occurs in two phases. The first phase identifies the most prevalent categories across a set of portals, while the second phase aligns the categories of individual portals with these overall prevalent categories.

By examining historical download rates and user interactions with datasets, our approach provides valuable insights that can guide proactive decision-making regarding data publication and prioritization. This study offers a structured methodology for identifying HVDs with significant reuse potential, as demonstrated through a case study involving U.S. municipalities. This methodology aids in data prioritization and supports open governance initiatives by revealing disparities and commonalities in dataset utilization. By enhancing localized understanding of HVD, the research informs efforts to improve data homogeneity and interoperability within specific regions. Additionally, the approach involves automated analysis and visualization of downloaded data from Open Government Data (OGD) portals, guiding the prioritization of crucial datasets for open governance and data reuse initiatives.

On the theoretical side, the study introduces the High-Value Data index (HVDi), a novel metric for assessing dataset categorization across multiple portals. This index facilitates meaningful comparisons and identification of HVD within specific areas, exemplified through the case of U.S. municipalities. The methodology for automated analysis of download data offers a new perspective on understanding region-specific HVD, thereby enriching the theoretical framework for data prioritization and open governance. This approach emphasizes the importance of region-specific insights, moving beyond generally acknowledged HVD to address more localized needs and priorities.

While our method provides valuable insights into category-level trends, integrating an ex-ante approach would enhance the precision of identifying individual datasets with high value, ultimately contributing to more informed and proactive decision-making regarding data publication and prioritization on OGD portals. Thus, the proposed method can be seen as a semi-ex-ante approach, rooted in ex-post assessments. While our methodology primarily relies on analyzing past data usage patterns (ex-post assessment), it also incorporates elements of anticipation and foresight (semi-ex-ante approach) by leveraging these insights to inform future decisions. However, to fully embrace an ex-ante approach, incorporating prospective considerations and predictive analytics would further enhance the method's ability to anticipate future trends and preemptively identify HVDs.


\bibliographystyle{cas-model2-names}

\bibliography{bibliography}

\appendix
\setcounter{table}{0}
\renewcommand{\thetable}{A\arabic{table}}

\section{Appendix}\label{appendix}
\input{appendix}

\end{document}

%% file: appendix.tex
\newpage

\begin{landscape}
{\tiny\tabcolsep=3pt  
\begin{longtable}{p{2.2cm}p{2cm}p{2cm}p{2cm}p{2cm}p{2cm}p{2cm}p{2cm}p{2cm}p{2cm} }
\caption{Portals coverage alignment per category}
\label{tab:portals_coverage_alignmnet}\\
\toprule
                  & \textbf{Austin}                          & \textbf{New York}               & \textbf{Buffalo}                              & \textbf{Chicago}                                & \textbf{Los Angeles}                               & \textbf{Dallas}        & \textbf{San Francisco}                         & \textbf{Seattle}                & \textbf{Honolulu}       
\\ \midrule
%
\textbf{Public Safety}          & Public Safety                   & Public Safety          & Public Safety                        & Public Safety                          & Public Safety                             & Public Safety & Public Safety                         & Public Safety          & Public   Safety \\
\textbf{Transportation}         & Transportation                  & Transportation         & Transportation                       & Transportation                         & Transportation                            &               & Transportation                        & Transportation         & Transportation  \\
\textbf{Recreation}             & Recreation and   Culture        & Recreation             & Recreation                           & Ethics                                 & Parks \&   Recreation                     &               & Culture and   Recreation              & Parks and   Recreation & Recreation      \\
\textbf{Environment}            & Environment                     & Environment            & Energy \& Environment                & Environment \& Sustainable Development & Sanitation                                &               & Energy and Environment                &                        &                 \\
\textbf{Finance}                & Agriculture                     &                        &                                      & Administration   \& Finance            & Administration   \& Finance               &               &                                       & Finance                & Finance         \\
\textbf{Economic   Development} & Building and Development        & Housing \& Development & Economic \& Neighborhood Development & Community \& Economic Development      & Community \& Economic Development         &               &                                       &                        &                 \\
\textbf{Infrastructure}         & City   Infrastructure           &                        & Infrastructure                       &                                        & City   Infrastructure \& Service Requests &               & City   Infrastructure                 &                        &                 \\
\textbf{Business}               & Business and Economy            & Business               &                                      &                                        &                                           &               &                                       & City Business          & Business        \\
\textbf{Boundaries}             &                                 &                        & Transparency                         & Facilities   \& Geographic Boundaries  &                                           &               & Geographic   Locations and Boundaries &                        & Location        \\
\textbf{Education}              &                                 & Education              & Education                            & Education                              & Education                                 &               &                                       &                        &                 \\
\textbf{City Services}          &                                 &                        & Human Services                       &                                        & A Livable and   Sustainable City          & City Services & City Management   and Ethics          &                        &                 \\
\textbf{Social   Services}      & Social Services                 & Social Services        &                                      &                                        &                                           & Services      & Health and Social Services            &                        &                 \\
\textbf{Community}              & Health and   Community Services &                        &                                      &                                        &                                           &               & Economy and   Community               & Community              &                 \\
\textbf{Health}                 &                                 & Health                 & Health                               & Health \& Human Services               &                                           &               &                                       &                        &                 \\
\textbf{Government}             & City Government                 & City Government        & Government                           &                                        &                                           &               &                                       &                        &                 \\
\textbf{Facilities}             & Locations and Maps              &                        &                                      & Buildings                              &                                           &               & Housing and Buildings                 &                        &                 \\
\textbf{Permits}                & Permits and   Licensing         &                        &                                      &                                        & Audits and   Reports                      &               &                                       &                        &                 \\
\textbf{Property}               & Housing and Real Estate         &                        &                                      &                                        & Housing and Real Estate                   &               &                                       &                        &                 \\
\textbf{Planning}               &                                 &                        &                                      &                                        & Aging                                     &               &                                       & Permitting             &                 \\
\textbf{Parks}                  &                                 &                        &                                      &                                        &                                           &               &                                       & Parks                  &                 \\
\textbf{Economy}                &                                 &                        &                                      &                                        &                                           & Economy       &                                       &                        &                 \\
\textbf{Culture}                &                                 &                        &                                      &                                        & Arts \& Culture                           &               &                                       &                        &                 \\
\textbf{Budget}                 &                                 &                        &                                      &                                        & Budget                                    &               &                                       &                        &                 \\
\textbf{Demographics}           &                                 &                        &                                      &                                        & Statistics                                &               &                                       &                        &                 \\
\textbf{Service Requests}       &                                 &                        &                                      & Service   Requests                     &                                           &               &                                       &                        &                 \\
\textbf{Utilities}              & Utilities and City Services     &                        &                                      &                                        &                                           &               &                                       &                        &                 \\
\textbf{Land Records}           &                                 &                        &                                      &                                        &                                           &               &                                       & Land Base              &                     \\ \bottomrule

\end{longtable}
 }

\end{landscape}